\documentclass[preprint,3p,12pt]{elsarticle}

\usepackage{amssymb}
\usepackage{amsmath}
\usepackage{booktabs}
\usepackage{nicefrac}
\usepackage{graphicx}
\usepackage{lscape}
\usepackage{multicol}
\usepackage{tabularray}
\usepackage{subcaption}
\usepackage{nicematrix}
\usepackage{gensymb}
\usepackage{makecell}
\usepackage{xcolor}
\makeatletter
\def\maketag@@@#1{\hbox{\m@th\normalfont\normalsize#1}}
\makeatother

\usepackage{lineno}

\usepackage{diagbox}
\usepackage[colorlinks,bookmarksopen,bookmarksnumbered,citecolor=red,urlcolor=red]{hyperref}

\biboptions{sort&compress}
\journal{Engineering Fracture Mechanics}

\usepackage{nomencl}
\renewcommand\nomgroup[1]{%
  \item[\marthbf{%
  \ifthenelse{\equal{#1}{A}}{Latin characters}{%
  \ifthenelse{\equal{#1}{B}}{Greek characters}{%
  \ifthenelse{\equal{#1}{C}}{Sub/superscripts}{%
  \ifthenelse{\equal{#1}{D}}{Abbreviations}{}}}}%
  }]\ignorespaces
}

\begin{document}
\newcommand{\I}{\rm{I}}
\newcommand{\II}{\rm{II}}
\newcommand{\lce}{l}
\newcommand{\beam}{\rm{beam}}
\newcommand{\btm}{\rm{bot}} 
\newcommand{\CE}{\rm{CE}}
\newcommand{\CR}{\rm{CR}}
\newcommand{\dd}{\mathrm{d}} 
\newcommand{\external}{\rm{ext}}
\newcommand{\h}{\rm{h}}
\newcommand{\internal}{\rm{int}}
\newcommand{\len}{\rm{\mathcal{l}}}
\newcommand{\truss}{\rm{truss}}
\newcommand{\tp}{\rm{top}} 
\newcommand{\transpose}{^{\rm{T}}}
\newcommand{\invtranspose}{^{\rm{-T}}}
\newcommand{\xl}{{\hat{x}}}
\newcommand{\yl}{{\hat{y}}}
\newcommand{\inv}{\makebox[0pt][l]{$^{-1}$}}
\newcommand{\DeltaI}{\Delta_{\text{I}}}
\newcommand{\DeltaII}{\Delta_{\text{II}}}
\newcommand{\DeltaIII}{\Delta_{\text{III}}}
\newcommand{\wtopce}{w^{\text{topCE}}}
\newcommand{\wbotce}{w^{\text{botCE}}}
\newcommand{\utopce}{u^{\text{topCE}}}
\newcommand{\ubotce}{u^{\text{botCE}}}
\newcommand{\vtopce}{v^{\text{topCE}}}
\newcommand{\vbotce}{v^{\text{botCE}}}
\newcommand{\wtop}{w^{\text{top}}}
\newcommand{\wbot}{w^{\text{bot}}}
\newcommand{\utop}{u^{\text{top}}}
\newcommand{\ubot}{u^{\text{bot}}}
\newcommand{\vttop}{v^{\text{top}}}
\newcommand{\vbot}{v^{\text{bot}}}
\newcommand{\htop}{h^{\text{top}}}
\newcommand{\hbot}{h^{\text{bot}}}
\newcommand{\thetatop}{\theta^{\text{top}}}
\newcommand{\thetabot}{\theta^{\text{bot}}}

\newcommand{\ba}{\boldsymbol{\mathrm{a}}}
\newcommand{\bA}{\boldsymbol{\mathrm{A}}}
\newcommand{\bb}{\boldsymbol{\mathrm{b}}}
\newcommand{\bB}{\boldsymbol{\mathrm{B}}}
\newcommand{\bBN}{\bB_{\text{N}}}
\newcommand{\bC}{\boldsymbol{\mathrm{C}}}
\newcommand{\D}{\boldsymbol{\mathrm{D}}}
\newcommand{\Dce}{\D^{\CE}}
\newcommand{\bE}{\boldsymbol{\mathrm{E}}}
\newcommand{\bF}{\boldsymbol{\mathrm{F}}}
\newcommand{\f}{\boldsymbol{\mathrm{f}}} 
\newcommand{\fext}{\f_{\external}}
\newcommand{\fint}{\f_{\internal}}
\newcommand{\fhatext}{\hat{\f}_{\external}}
\newcommand{\bH}{\boldsymbol{\mathrm{H}}}
\newcommand{\bI}{\boldsymbol{\mathrm{I}}}
\newcommand{\bJ}{\boldsymbol{\mathrm{J}}}
\newcommand{\K}{\boldsymbol{\mathrm{K}}}
\newcommand{\Kmat}{\K_{\mathrm{mat}}}
\newcommand{\Kgeo}{\K_{\mathrm{geo}}}
\newcommand{\bM}{\boldsymbol{\mathrm{M}}}
\newcommand{\bn}{\boldsymbol{\mathrm{n}}}
\newcommand{\bN}{\boldsymbol{\mathrm{N}}}
\newcommand{\bp}{\boldsymbol{\mathrm{p}}}
\newcommand{\bq}{\boldsymbol{\mathrm{q}}}
\newcommand{\bqce}{\bq^{\CE}}
\newcommand{\bQ}{\boldsymbol{\mathrm{Q}}}
\newcommand{\bU}{\boldsymbol{\mathrm{U}}}
\newcommand{\bR}{\boldsymbol{\mathrm{R}}}
\newcommand{\bS}{\boldsymbol{\mathrm{S}}}
\newcommand{\bt}{\boldsymbol{\mathrm{t}}}
\newcommand{\bT}{\boldsymbol{\mathrm{T}}}
\newcommand{\bu}{\boldsymbol{\mathrm{u}}}
\newcommand{\bv}{\boldsymbol{\mathrm{v}}}
\newcommand{\bw}{\boldsymbol{\mathrm{w}}}
\newcommand{\bW}{\boldsymbol{\mathrm{W}}}
\newcommand{\bX}{\boldsymbol{\mathrm{X}}}
\newcommand{\bx}{\boldsymbol{\mathrm{x}}}

\newcommand{\balpha}{\boldsymbol{\mathrm{\alpha}}}
\newcommand{\bDelta}{\boldsymbol{\mathrm{\Delta}}}
\newcommand{\beps}{\boldsymbol{\mathrm{\epsilon}}}
\newcommand{\bgamma}{\boldsymbol{\mathrm{\gamma}}}
\newcommand{\bPhi}{\boldsymbol{\Phi}}
\newcommand{\bsig}{\boldsymbol{\mathrm{\sigma}}}
\newcommand{\btau}{\boldsymbol{\mathrm{\tau}}}
\newcommand{\bxi}{\boldsymbol{\mathrm{\xi}}}

\newcommand{\minus}{\scalebox{0.75}[1.0]{$-$}}
\begin{frontmatter}



\title{Accurate simulation of delamination with a resin-rich layer-dependent penalty stiffness based on structural cohesive elements}


\author[inst1]{Xiaopeng Ai} 

\affiliation[inst1]{organization={Department of Aerospace Structures and Materials, Faculty of Aerospace Engineering, Delft University of Technology},
            addressline={Kluyverweg 1}, 
            city={Delft},
            postcode={2629 HS}, 
            country={The Netherlands}}
            
\affiliation[inst2]{organization={School of Optoelectronic Science and Engineering, Soochow University},
            addressline={Shizi Street 1}, 
            city={Suzhou},
            postcode={215031}, 
            country={China}}

\author[inst1]{Christos Kassapoglou}
\author[inst2,inst1]{Boyang Chen\corref{cor1}}
 \ead{boyangchen@suda.edu.cn}
 \cortext[cor1]{Corresponding author}

\begin{abstract}
Shell-based cohesive elements tend to overestimate the compression ahead of the crack tip because of approximations in the penalty stiffness. In this study, the higher-order structural cohesive element previously developed by the authors is enhanced with a resin-rich layer-dependent penalty stiffness to improve both computational efficiency and predictive accuracy. The proposed formulation distinguishes between the normal and shear penalty stiffnesses. It extends the resin-rich layer-based penalty stiffness from the layer-wise to the equivalent single-layer framework. This extension is achieved using the through-thickness distributions of the out-of-plane normal and transverse shear stresses derived from beam theory. The proposed method is verified and validated against benchmark problems for Mode I, Mode II, mixed-mode, and reinforced DCB configurations. Compared with the conventional formulation, it exhibits significantly improved mesh convergence and provides more accurate predictions of the compression distribution ahead of the crack tip and the delamination propagation.
\end{abstract}

\begin{graphicalabstract}
\end{graphicalabstract}

\begin{highlights}
\item A new penalty stiffness formulation for structural cohesive elements is proposed.
\item The formulation accounts for out-of-plane normal and transverse shear stresses through the thickness.
\item Resin-rich layer properties are incorporated into the penalty stiffness derivation.
\item The formulation improves mesh convergence and prediction accuracy.
\item The method is validated by Mode I, Mode II, mixed-mode, and R-DCB benchmarks.
\end{highlights}

\begin{keyword}
Composites laminates \sep Delamination \sep Cohesive element  \sep Penalty stiffness


\end{keyword}

\end{frontmatter}

\makenomenclature
\nomenclature[A, 01]{\(a_0\)}{pre-crack length}
\nomenclature[A, 01]{\(a_1,a_2,a_3,a_4\)}{unknown coefficients of the stress distribution}
\nomenclature[A, 02]{\(B\)}{specimen width}

\nomenclature[A, 02]{\(E_3\)}{out-of-plane elastic modulus of the laminate}
\nomenclature[A, 02]{\(E_{\textrm{rr}}\)}{elastic modulus of the resin-rich layer}

\nomenclature[A, 02]{\(G_{13}\)}{transverse shear modulus of the laminate}
\nomenclature[A, 02]{\(G_{\textrm{II}}\)}{Mode II energy release rate}
\nomenclature[A, 02]{\(G_{\textrm{T}}\)}{total energy release rate}
\nomenclature[A, 02]{\(G_{\textrm{rr}}\)}{shear modulus of the resin-rich layer}

\nomenclature[A, 02]{\(h\)}{thickness of the laminate}
\nomenclature[A, 02]{\(h_{\textrm{bot}}\)}{thickness of the bottom arm of the specimen}
\nomenclature[A, 02]{\(h_{\textrm{top}}\)}{thickness of the top arm of the specimen}
\nomenclature[A, 02]{\(h_{\textrm{rr}}\)}{thickness of the resin-rich layer}

\nomenclature[A, 03]{\(K_n\)}{normal penalty stiffness}
\nomenclature[A, 03]{\(K_s\)}{shear penalty stiffness}
\nomenclature[A, 03]{\(K_n^{\textrm{Ai}}\)}{normal penalty stiffness proposed by Ai et al.}
\nomenclature[A, 03]{\(K_s^{\textrm{Ai}}\)}{shear penalty stiffness proposed by Ai et al.}
\nomenclature[A, 03]{\(K_s^{\textrm{Bazilevs}}\)}{shear penalty stiffness proposed by Bazilevs et al.}
\nomenclature[A, 03]{\(K^{\textrm{Turon}}\)}{penalty stiffness proposed by Turon et al.}

\nomenclature[A, 03]{\(L\)}{half length of the specimen}

\nomenclature[A, 03]{\(n_{\textrm{top}}, n_{\textrm{bot}}\)}{the number of plies in the top and bottom arms of the specimen}
\nomenclature[A, 03]{\(n_{\textrm{total}}\)}{the total number of plies}

\nomenclature[B,01]{\(\alpha\)}{scaling parameter}
\nomenclature[B,02]{\(\Delta_{\textrm{I}},\,\Delta_{\textrm{II}}\)}{Mode I and Mode II opening displacements}
\nomenclature[B,03]{\(\varepsilon_z\)}{out-of-plane normal strain}
\nomenclature[B,04]{\(\varepsilon_{xz}\)}{transverse shear strain}
\nomenclature[B,05]{\(\nu_{\textrm{rr}}\)}{Poisson's ratio of the resin-rich layer}
\nomenclature[B,05]{\(\sigma_z\)}{out-of-plane normal stress}
\nomenclature[B,06]{\(\sigma_z^{\textrm{max}}\)}{maximum out-of-plane normal stress}
\nomenclature[B,07]{\(\sigma_z^{\textrm{top}}\)}{out-of-plane normal stress in the top arm of the specimen}
\nomenclature[B,08]{\(\sigma_z^{\textrm{bot}}\)}{out-of-plane normal stress in the bottom arm of the specimen}
\nomenclature[B,09]{\(\tau_{xz}\)}{transverse shear stress}
\nomenclature[B,10]{\(\tau_{xz}^{\textrm{max}}\)}{maximum transverse shear stress}
\nomenclature[B,11]{\(\tau_{xz}^{\textrm{top}}\)}{transverse shear stress in the top arm of the specimen}
\nomenclature[B,12]{\(\tau_{xz}^{\textrm{bot}}\)}{transverse shear stress in the bottom arm of the specimen}

\nomenclature[C,03]{\(\bullet^{\mathrm{Mode~I}}\)}{related to pure Mode I loading}
\nomenclature[C,03]{\(\bullet^{\mathrm{Mode~II}}\)}{related to pure Mode II loading}
\nomenclature[C,03]{\(\bullet_n\)}{related to the normal direction}
\nomenclature[C,04]{\(\bullet_s\)}{related to the shear direction}
\nomenclature[C,01]{\(\bullet^\mathrm{DCB}\)}{related to the DCB specimen}
\nomenclature[C,02]{\(\bullet^\mathrm{ENF}\)}{related to the ENF specimen}
\nomenclature[C,03]{\(\bullet^\mathrm{MMB}\)}{related to the MMB specimen}
\nomenclature[C,08]{\(\bullet_z\)}{related to the global \(z\)-direction}
\nomenclature[C,09]{\(\bullet_{xz}\)}{related to the global \(xz\)-plane}
\nomenclature[C,07]{\(\bullet^\mathrm{Unreinforced}\)}{related to the unreinforced region of the R-DCB specimen}
\nomenclature[C,07]{\(\bullet^\mathrm{Reinforced}\)}{related to the reinforced region of the R-DCB specimen}
\nomenclature[C,07]{\(\bullet^\mathrm{Undamaged}\)}{related to the undamaged region of the R-DCB specimen}

\nomenclature[D, 01]{CE}{Cohesive Element}
\nomenclature[D, 01]{CZL}{Cohesive Zone Length}
\nomenclature[D, 01]{DCB}{Double Cantilever Beam}
\nomenclature[D, 01]{ENF}{End-Notched-Flexure}
\nomenclature[D, 01]{MMB}{Mixed-Mode Bending}
\nomenclature[D, 01]{FRP}{Fiber-Reinforced Polymer}
\nomenclature[D, 01]{R-DCB}{Reinforced Double Cantilever Beam}

\printnomenclature[0.8in] 


\section{Introduction}
\label{sec:Introduction}
Fiber-reinforced polymer (FRP) composites are widely used in aerospace structures~\cite{baker2004composite}. The prediction of strength and damage tolerance is an indispensable aspect in the design of composites
aircraft structures~\cite{kassapoglou2010design,chen2013numerical}. Among the various failure mechanisms, delamination is one principal mode of composite failure that occurs between the layers in the composite laminate~\cite{ratcliffe2013test,chen2017modelling}. Progressive failure prediction during delamination is essential because reliable numerical simulations can reduce the need for expensive physical experiments under different loading conditions. The cohesive element (CE)~\cite{dugdale1960yielding,barenblatt1962mathematical}, based on the cohesive zone model (CZM), is widely employed as an interface element to simulate delamination by capturing the fracture process zone ahead of the crack tip.
However, the standard CE is subject to a well-known limitation related to mesh density: the element size must be smaller than the cohesive zone length (CZL) to ensure accurate prediction of delamination because of high-stress gradients at the crack tip. 

Recently, a higher-order structural cohesive element method has been developed to model delamination of composite laminates based on the Kirchhoff-Love hypothesis~\cite{ai2025structural}. This approach overcomes the cohesive zone limitation while significantly improving computational efficiency by replacing solid hexahedral elements with higher-order triangular shell elements.
The primary objective of that study was to develop a higher-order structural cohesive element. Consequently, the formulation focused mainly on the strain-displacement matrix (B-matrix), whereas the constitutive matrix (D-matrix) received little attention. A conventional cohesive law proposed by Turon et al.~\cite{turon2006damage,turon2007engineering} was adopted as the constitutive model, in which the penalty stiffness is defined by the laminate thickness and its out-of-plane elastic modulus. Although this formulation has been widely adopted for both solid~\cite{turon2010accurate,lu2019cohesive} and shell elements~\cite{davila2007cohesive,davila2008effective,balducci2024overcoming}, the penalty stiffness significantly influences the prediction of through-thickness stress distributions in Kirchhoff-Love shell models.
Since these stresses play an important role in the initiation of in-plane failure, such as matrix cracking, the accurate prediction of these stresses is particularly important. Previous studies on two- and three-dimensional structural cohesive elements based on Kirchhoff--Love theory showed that the conventional penalty stiffness failed to accurately predict the normal traction distribution ahead of the crack tip~\cite{russo2020overcoming,balducci2024overcoming}. In particular, the compression was significantly overestimated, and mesh convergence could not be achieved.
Several researchers have proposed alternative penalty stiffness formulations for thin-shell elements based on the laminate thickness and material properties. \citet{polla2021delamination} proposed a method to evaluate both normal and shear penalty stiffness. In this formulation, the cohesive element thickness is defined as the sum of the thicknesses of the top and bottom shell elements. The formulation was evaluated against Turon's penalty stiffness in the DCB and ENF benchmarks, but no significant improvement was observed. The study reported only the load-displacement curves of the benchmark problems, without providing any analysis of the traction distribution.
\citet{bazilevs2018new} proposed a method for determining shear penalty stiffness by defining the thickness of the cohesive element as half the sum of the thicknesses of the adjacent shell elements. The formulation was validated on both the DCB and ENF benchmark models. However, the ENF results showed a significant underestimation of initial stiffness and the peak load compared with the analytical solution. Despite their different definitions of the cohesive element thickness, both Bazilevs' and Polla's formulations assume simple shear deformation of the plies under transverse shear loading.

Ai et~al.~\cite{ai2026newapproachdeterminationthroughthickness} proposed a resin-rich-layer-based formulation that accurately predicts the through-thickness stress distribution in multilayer composite laminates.
The formulation has been successfully applied to higher-order structural elements, leading to a significant improvement in computational efficiency. The penalty stiffness obtained by this formulation only depends on the thickness and material properties of the resin-rich layer between adjacent plies. It requires layer-wise modeling of multi-layer composite laminates. This approach is acceptable for laminates with a relatively small number of plies. However, for laminates containing many plies with the same fiber orientation, the model becomes computationally inefficient. These plies are more efficiently represented as a single ply block using one shell element. Consequently, the formulation cannot be directly applied to equivalent single-layer models.

To overcome this limitation, the present study develops a new penalty stiffness formulation for structural cohesive elements used between shell elements representing multi-ply blocks. The proposed formulation preserves the accuracy of the resin-rich-layer approach while being directly applicable to equivalent single-layer models. The structure of the paper is organised as follows. Section~\ref{sec:method} presents the proposed penalty stiffness formulation. Section~\ref{sec:validation} validates the performance of the proposed formulation on a series of benchmarks on delamination. Section~\ref{sec:summary} summarizes the conclusions of this study and outlines potential directions for future research.
\section{Method}\label{sec:method}
\subsection{Cubic structural element}
\begin{figure}[h!]
	\centering
	\includegraphics[width=0.8\linewidth]{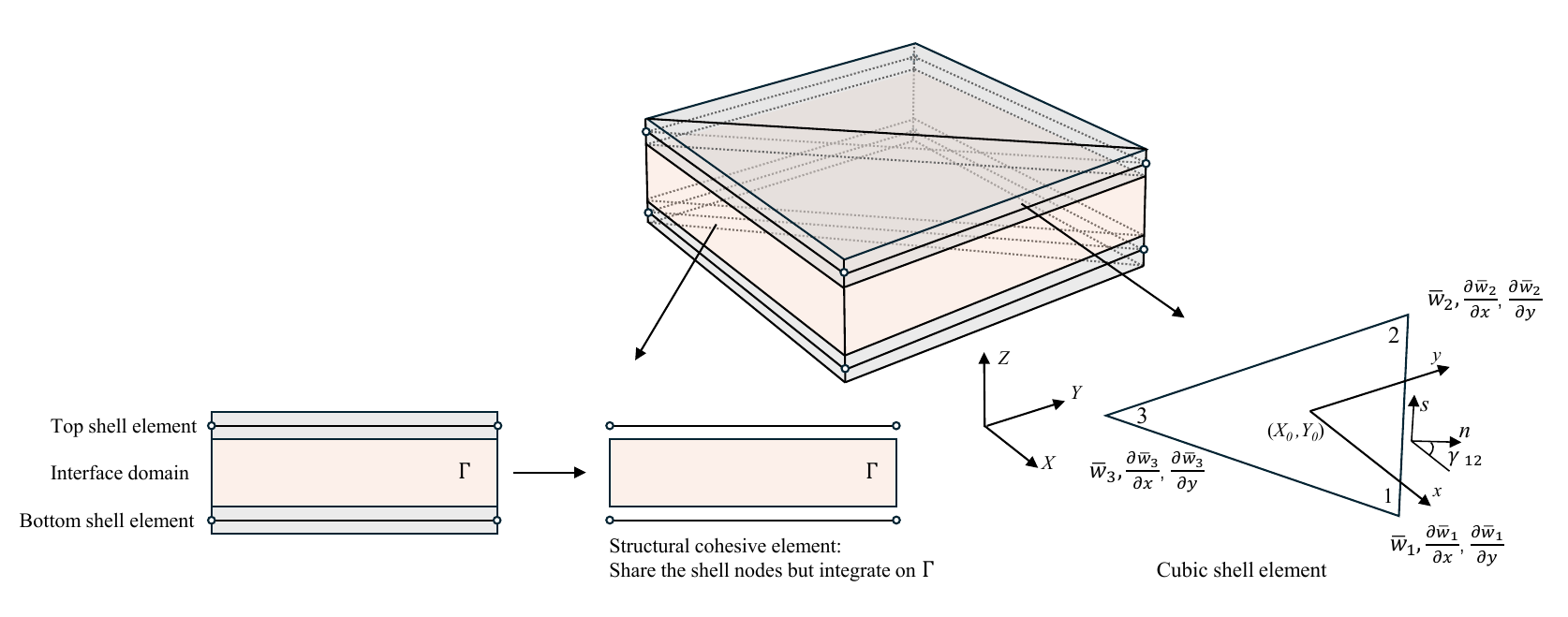}
	\caption{Triangular cubic structural element~\cite{ai2025structural}}
	\label{fig:Structural element}
\end{figure}

In this work, all benchmark models are discretized by high-order triangular structural elements, which are implemented through user-defined element subroutines in Abaqus. Cubic shell elements represent the laminate plies, while the interfaces between the adjacent shell elements are represented by compatible cohesive elements, as illustrated in Figure~\ref{fig:Structural element}. This element overcomes the cohesive zone limitation, while the use of triangular elements enhances geometric flexibility. The detailed kinematic description and stiffness matrix formulation of the element can be found in Reference~\cite{ai2025structural}.

\subsection{Penalty stiffness based on the resin-rich layer}
\begin{figure}[h!]
      \centering
	   \begin{subfigure}{0.46\linewidth}
		\includegraphics[width=\linewidth]{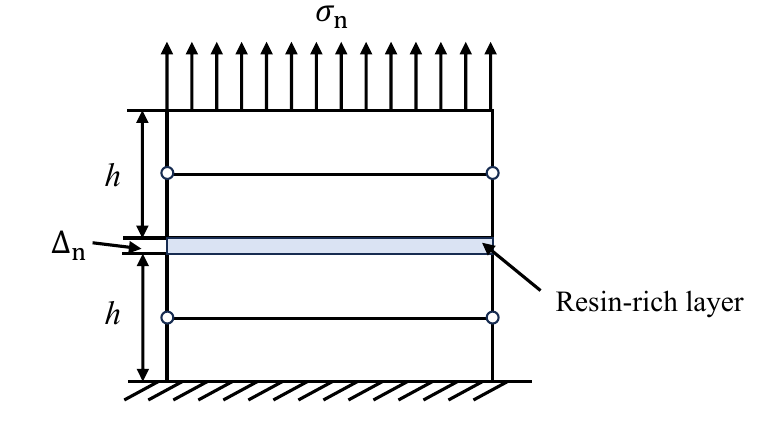}
		\caption{Normal penalty stiffness}
		\label{fig:Kn-2ndPaper}
	   \end{subfigure}
	   \begin{subfigure}{0.46\linewidth}
		\includegraphics[width=\linewidth]{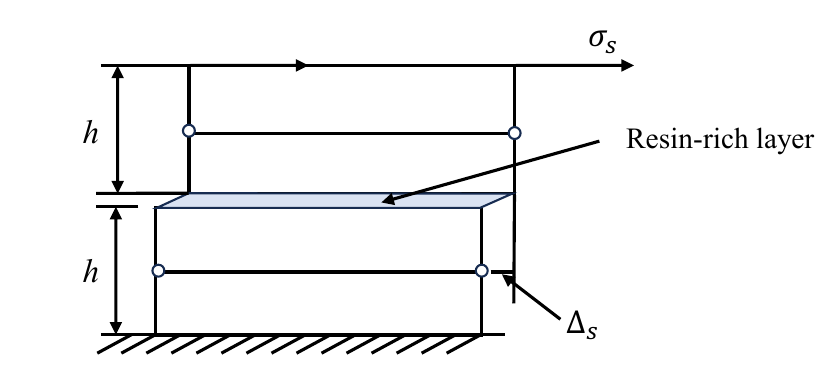}
		\caption{Shear penalty stiffness}
		\label{fig:Ks-2ndPaper}
	    \end{subfigure}
	\caption{Resin-rich layer-based penalty stiffness~\cite{ai2026newapproachdeterminationthroughthickness}}
	\label{fig:K-2ndPaper}
\end{figure}

The conventional penalty stiffness formulation for cohesive elements, mentioned in Section~\ref{sec:Introduction}, is proposed by Turon et al.~\cite{turon2007engineering}:
\begin{align}
   K^{\textrm{Turon}}= \alpha \,\frac{E_3}{h}
\end{align} 
where $E_3$ is the laminate out-of-plane elastic modulus and $h$ is the laminate thickness. The parameter $\alpha$ is a scaling parameter which is typically chosen sufficiently large to prevent artificial separation before damage initiation. The excessively large penalty stiffness value may lead to numerical instability and convergence difficulties. When this formulation is applied to shell-based cohesive elements, the out-of-plane stress distribution through the thickness cannot be accurately predicted. To overcome this limitation, an accurate penalty stiffness was derived based on a resin-rich layer in our previous study~\cite{ai2026newapproachdeterminationthroughthickness}. The normal and shear penalty stiffnesses were derived separately under the pure tension and simple shear loading conditions, as shown in Figure~\ref{fig:K-2ndPaper}:
\begin{align}\label{eq:K-2ndPaper}
    K^{\textrm{Ai}}_n &= \frac{E_{\textrm{rr}}}{h_{\text{rr}}} \\[0.5em]
    K^{\textrm{Ai}}_s &=  \frac{G_{\textrm{rr}}}{h_{\text{rr}}}
\end{align}
where $E_{\textrm{rr}}$ and $G_{\textrm{rr}}$ are the elastic modulus and shear modulus of the resin-rich layer material, respectively, and $h_{\text{rr}}$ is the thickness of the resin-rich layer at the ply interface of interest. This formulation provides a clear physical interpretation as the penalty stiffness directly corresponds to the transverse stiffness of the resin-rich layer. The formulation assumes that the interlaminar normal and shear stresses are transferred through the resin-rich layer, which is significantly more compliant than the adjacent composite plies. Although this method accurately predicts the through-thickness stress distribution, it requires structural cohesive elements to be inserted between every pair of adjacent plies, regardless of whether the plies have the same fiber orientations.

In many practical applications, such as the DCB benchmark~\cite{borg2004simulating,li2014combined}, adjacent plies with the same fiber orientation are homogenized into a single shell element using the equivalent single-layer approach. This method significantly reduces the number of finite elements and thereby improves computational efficiency. When the equivalent single-layer approach is employed, the parameter $h$ in Figure~\ref{fig:K-2ndPaper} no longer represents the thickness of an individual ply but that of the entire ply block. Therefore, the interface penalty stiffness should account for the transverse stiffness of all the resin-rich layers contained within the homogenized laminate. To achieve this objective, the through-thickness stress distribution is incorporated while preserving the original physical meaning of $h_{\textrm{rr}}$, thereby enabling the determination of an accurate penalty stiffness. The derivation of a physically reasonable through-thickness stress distribution is a key component of the proposed method. The derivations of the corresponding normal and shear penalty stiffnesses are presented in Section~\ref{subsubsec:shear_K}.
Because the out-of-plane normal and transverse shear stress distributions exhibit distinct characteristics, separate normal and shear penalty stiffnesses are derived. Accordingly, the mode-dependent cohesive law proposed by Turon et al.~\cite{turon2010accurate} is adopted in the structural cohesive element formulation.

\subsection{Derivation of the normal and shear penalty stiffnesses}\label{subsubsec:shear_K}
Since Kirchhoff-Love shell elements do not explicitly account for transverse stresses, it is necessary to determine the through-thickness stress distributions based on classical mechanics. To facilitate the derivation, the three-dimensional shell problem is reduced to a two-dimensional beam problem. Under the Kirchhoff-Love assumptions, the axial stress varies linearly through the thickness, whereas the transverse shear stress follows a quadratic distribution. For consistency with the laminate coordinate system adopted in this work, the transverse coordinate in the beam formulation is hereafter represented by $z$. Accordingly, the transverse shear stress is assumed to take the following form: 
\begin{align}\label{eq:tau_xy-2nd} 
    \tau_{xz} =f(x)* (a_1z^2+a_2z+a_3)
\end{align}
where $f(x)$ describes the variation along $x$ and $a_1$, $a_2$, and $a_3$ are unknown coefficients

The corresponding two-dimensional equilibrium equation is:
\begin{align}
\frac{\partial \tau_{xz}}{\partial x} + \frac{\partial \sigma_z}{\partial z} = 0 \label{eq:2d-equilibrium-2}
\end{align}

According to Equation~\ref{eq:2d-equilibrium-2}, a quadratic distribution of $\tau_{xz}$ leads to a cubic distribution of the out-of-plane normal stress $\sigma_z$, which can be expressed as
\begin{align}\label{eq:sigma_y-3rd} 
    \sigma_{z} = f^\prime(x)\left(\frac{a_1}{3}z^3+\frac{a_2}{2}z^2+a_3z+a_4\right)
\end{align}
where $a_4$ is an additional unknown coefficient.

The derivation of $K_n$ begins with the through-thickness distribution of the out-of-plane normal stress. According to Hooke's law, the out-of-plane normal strain at the $i^{\textrm{th}}$ interface can be obtained:
\begin{align}\label{eq:normal_stress-normal_strain}
    \varepsilon_{z}(z_i) = \frac{\sigma_z(z_i)}{E_{\textrm{rr}} } 
\end{align}
where $z_i$ represents the coordinate of the $i^{\textrm{th}}$ interface layer. The $\sigma_z(z_i)$ and $\varepsilon_{z}(z_i)$ represent the out-of-plane normal stress and strain values at the $i^{\textrm{th}}$ interface layer. $E_{\textrm{rr}}$ is the elastic modulus of the resin-rich layer.

For pure Mode I loading, i.e., the Double Cantilever Beam (DCB) model, the top and bottom arms are subjected to different boundary conditions. Consequently, the contributions of the two arms must be considered separately. The total opening displacement of the cohesive element can  be expressed as the sum of the product of normal strains at each interface and the physical thickness of the resin-rich layer $h_{\textrm{rr}}$: 
\begin{align}\label{eq:DeltaI_total}
    \Delta^{\textrm{total}}_{\textrm{I}} &= \sum_{i=1}^{n_{\textrm{top}}}\, \varepsilon_{z}(z_i)\, h_{\textrm{rr}} +  \sum_{j=1}^{n_{\textrm{bot}}}\, \varepsilon_{z}(z_j)\, h_{\textrm{rr}} \nonumber\\[0.5em]
    &= \sum_{i=1}^{n_{\textrm{top}}}\, \frac{\sigma_z(z_i)}{E_{\textrm{rr}}}\, h_{\textrm{rr}} +  \sum_{j=1}^{n_{\textrm{bot}}}\, \frac{\sigma_z(z_j)}{E_{\textrm{rr}}}\, h_{\textrm{rr}}
\end{align}
where $n_{\textrm{top}}$ and $n_{\textrm{bot}}$ represent the number of plies in the top and bottom arms, respectively. The total number of plies $n_{\textrm{total}}$ is:
\begin{align}
    n_{\textrm{total}} = n_{\textrm{top}} + n_{\textrm{bot}}
\end{align}

Based on the traction-separation law of the cohesive model, the maximum out-of-plane normal stress can be expressed as:
\begin{align}\label{eq:sigma_max}
   \sigma_z^{\textrm{max}} = K_{n} \, \Delta^{\textrm{total}}_{\textrm{I}} 
\end{align}
where $\sigma_z^{\textrm{max}}$ is the maximum value of the out-of-plane normal stress distribution and $K_{n}$ is the normal penalty stiffness.

Substituting Equations~\ref{eq:normal_stress-normal_strain} and \ref{eq:DeltaI_total} into Equation~\ref{eq:sigma_max} leads to the following expression for the normal penalty stiffness:
\begin{align}\label{eq:Kn_final}
K_{n} &= \frac{\sigma_z^{\textrm{max}}}
{\displaystyle\sum_{i=1}^{n_{\textrm{top}}}\, \frac{\sigma_z(z_i)}{E_{\textrm{rr}}}\, h_{\textrm{rr}} +  \sum_{j=1}^{n_{\textrm{bot}}}\, \frac{\sigma_z(z_j)}{E_{\textrm{rr}}}\, h_{\textrm{rr}}} \nonumber\\[0.5em]
& = \frac{1}
{\displaystyle\sum_{i=1}^{n_{\textrm{top}}}\, \frac{\sigma_z(z_i)}{\sigma_z^{\textrm{max}}}\, \frac{h_{\textrm{rr}}}{E_{\textrm{rr}}} +  \sum_{j=1}^{n_{\textrm{bot}}}\, \frac{\sigma_z(z_j)}{\sigma_z^{\textrm{max}}}\,\frac{h_{\textrm{rr}}}{E_{\textrm{rr}}}}
\end{align}

Since $h_{\textrm{rr}}$ and $E_{\textrm{rr}}$ are geometric and material properties of the resin-rich layer, the quantity $h_{\textrm{rr}}/E_{\textrm{rr}}$ is known. Therefore, determining $K_{\textrm{n}}$ reduces to deriving the through-thickness distribution of the out-of-plane normal stress and, in particular, the stress ratios $\sigma_z(z_i)/\sigma_z^{\textrm{max}}$ shown in Equation~\ref{eq:Kn_final}.

The DCB model considered in this work is symmetric about the mid-plane. Therefore, the out-of-plane normal stress reaches its maximum at the mid-plane and vanishes at the upper and lower free surfaces:
\begin{align}
    &\sigma_z(0) = \sigma_z^{\textrm{max}}  \label{eq:BC-DCB-MID}\\[0.5em]
    &\sigma_z(\pm h) = 0  \label{eq:BC-DCB-TOP}
\end{align}
where $h$ is the thickness of each laminate arm.

Because the top and bottom arms of the DCB model are subjected to different boundary conditions, the out-of-plane normal stress distribution is derived separately for each arm. Accordingly, the cubic function given in Equation~\ref{eq:sigma_y-3rd} is applied separately to the top and bottom arms. For the top arm, the unknown coefficient $a_4$ can be derived from the boundary condition in Equation~\ref{eq:BC-DCB-MID}:
\begin{align}
   a_4 = \sigma_z^{\textrm{max}}
\end{align}

To simplify the derivation, the maximum out-of-plane normal stress is normalized to unity, i.e.
$\sigma_z^{\textrm{max}}=1$. Accordingly, $a_4 = 1$. By applying the free-surface boundary condition in Equation~\ref{eq:BC-DCB-TOP}, the following constraint on the unknown coefficients is obtained:
\begin{align}
    \frac{a_1}{3}h^3+\frac{a_2}{2}h^2+a_3h+1 = 0 \label{eq:coefficient-a1-a2-a3}
\end{align}

To determine the remaining unknown coefficients, the boundary conditions for the shear stress in the DCB model must also be considered. In contrast to the out-of-plane normal stress, the shear stress vanishes not only at the free surfaces but also at the mid-plane:
\begin{align}
    &\tau_{xz}(0) = 0  \label{eq:BC-DCB-MID-shear}\\[0.5em]
    &\tau_{xz}(\pm h) = 0  \label{eq:BC-DCB-TOP-shear}
\end{align}

Substituting the boundary conditions in Equations~\ref{eq:BC-DCB-MID-shear} and \ref{eq:BC-DCB-TOP-shear} into Equation~\ref{eq:tau_xy-2nd}, the following constraints on the unknown coefficients can be given as:     
\begin{align}
    &a_3 = 0 \label{eq:coefficient-a3} \\[0.5em]
    &a_1h^2+a_2h = 0\label{eq:coefficient-a1-a2}
\end{align}

Since the shear stress distribution in the top arm is positive over the interval $0<z<h$, the coefficient $a_1$ must be positive, i.e., $a_1 > 0$. Combining Equations~\ref{eq:coefficient-a3} and \ref{eq:coefficient-a1-a2} gives
\begin{align}
    a_1 &= \frac{6}{h^3}\\[0.5em]
    a_2 &= -\frac{6}{h^2}
\end{align}

Substituting the obtained coefficients into Equations~\ref{eq:tau_xy-2nd} and \ref{eq:sigma_y-3rd}, the following expressions for the shear stress and out-of-plane normal stress in the top arm are given by:
\begin{align}
    \tau_{xz}^{\textrm{top}} &\propto\frac{6}{h^3}z^2-\frac{6}{h^2}z\label{eq:tau_xy-DCB-top}  \\[0.5em]
    \sigma_{z}^{\textrm{top}} &\propto \frac{2}{h^3}z^3-\frac{3}{h^2}z^2+1 \label{eq:sigma_y-DCB-top}
\end{align}

The stress distributions in the bottom arm can be derived using the same procedure. The unknown coefficients are determined by enforcing the boundary conditions given in Equations~\ref{eq:BC-DCB-MID}, \ref{eq:BC-DCB-TOP}, \ref{eq:BC-DCB-MID-shear} and \ref{eq:BC-DCB-TOP-shear}. In addition, the out-of-plane normal stress distribution is required to remain symmetric about the laminate mid-plane. The corresponding coefficients are:
\begin{align}
  \text{DCB bottom arm:}\qquad\left\{
  \begin{aligned}
   a_1 &= -\frac{6}{h^3} \\[0.5em]  
   a_2 &= -\frac{6}{h^2} \\[0.5em] 
   a_3 &= 0 \\[0.5em] 
   a_4 &= 1
   \end{aligned}
   \right.
\end{align}

Substituting these coefficients into Equations~\ref{eq:tau_xy-2nd} and \ref{eq:sigma_y-3rd}, the expressions for the shear stress and out-of-plane normal stress distributions in the bottom arm are given as follows:
\begin{align}
    \tau_{xz}^{\textrm{bot}} &\propto-\frac{6}{h^3}z^2-\frac{6}{h^2}z\label{eq:tau_xy-DCB-bot}  \\[0.5em]
    \sigma_{z}^{\textrm{bot}} &\propto -\frac{2}{h^3}z^3-\frac{3}{h^2}z^2+1 \label{eq:sigma_y-DCB-bot}
\end{align}

Combining Equations~\ref{eq:tau_xy-DCB-top} and \ref{eq:tau_xy-DCB-bot} gives the piecewise expression for the shear stress distribution under pure Mode I opening:
\begin{align}\label{eq:tau_xy-DCB-Final}
    \tau_{xz}^{\textrm{Mode I}} \propto \begin{cases}
   \dfrac{6}{h^3}z^2-\dfrac{6}{h^2}z , & h\ge z \ge 0 \\[1em]
   -\dfrac{6}{h^3}z^2-\dfrac{6}{h^2}z  , & 0> z \ge-h
\end{cases}
\end{align}

Similarly, combining Equations~\ref{eq:sigma_y-DCB-top} and \ref{eq:sigma_y-DCB-bot} yields the following piecewise expression for the out-of-plane normal stress distribution:
\begin{align}\label{eq:sigma_y-DCB-Final}
    \sigma_{z}^{\textrm{Mode I}} \propto \begin{cases}
   \dfrac{2}{h^3}z^3-\dfrac{3}{h^2}z^2+1, & h\ge z \ge 0 \\[1em]
   -\dfrac{2}{h^3}z^3-\dfrac{3}{h^2}z^2+1   , & 0> z \ge-h
\end{cases}
\end{align}

\begin{figure}[h!]
	\centering
	\includegraphics[width=1\linewidth]{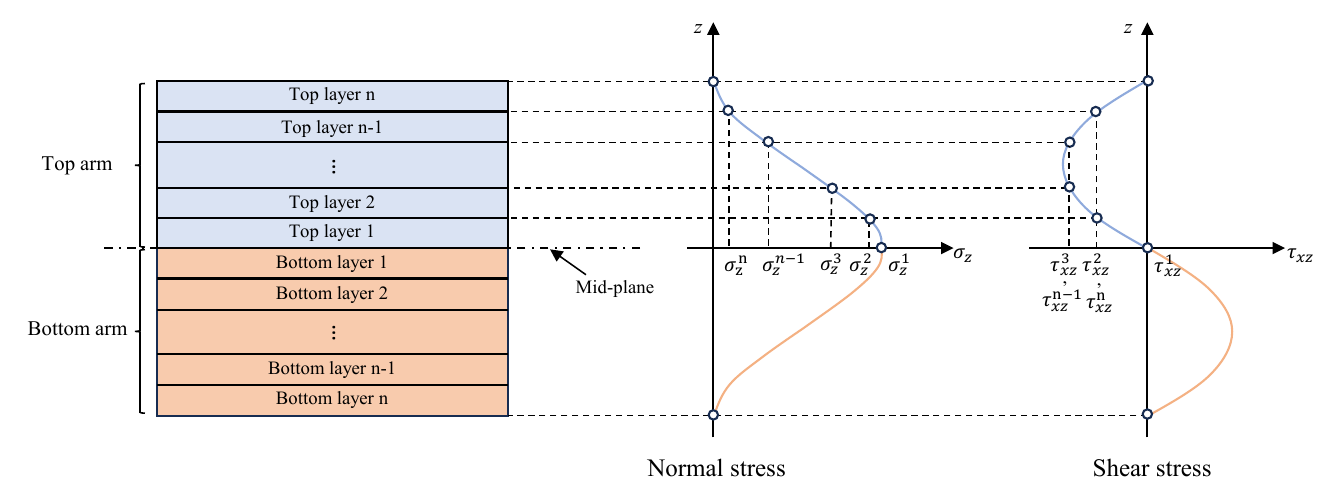}
	\caption{Stress distributions in the laminate under pure Mode I loading}
	\label{fig:SigmaY-TauXY-distribution-DCB}
\end{figure}

The through-thickness distributions of the shear stress and out-of-plane normal stress under pure Mode I loading are shown in Figure~\ref{fig:SigmaY-TauXY-distribution-DCB}. In the figure, $\sigma_z^i$ and $\tau_{xz}^i$ are the values of $\sigma_z^{\textrm{Mode I}}$ and $\tau_{xz}^{\textrm{Mode I}}$, respectively, at the $i^{\textrm{th}}$ interface layer. Once the through-thickness distribution of $\sigma_z^{\textrm{Mode I}}$ has been determined, the stress ratios $\sigma_z(z_i)/\sigma_z^{\textrm{max}}$ in Equation~\ref{eq:Kn_final} can be easily evaluated. Therefore, the normal penalty stiffness $K_n$ can be calculated directly.

For the shear penalty stiffness, the derivation is based on the through-thickness distribution of transverse shear stress, $\tau_{xz}$. According to Hooke's law, the corresponding shear strain at each interface can be obtained:
\begin{align}\label{eq:shear_stress-shear_strain}
    \varepsilon_{xz}(z_i) = \frac{\tau_{xz}(z_i)}{G_{\textrm{rr}} } 
\end{align}
where $\tau_{xz}(z_i)$ and $\varepsilon_{xz}(z_i)$ represent the transverse shear stress and strain at the $i^{\textrm{th}}$ interface layer, respectively. In addition, $G_{\textrm{rr}}$ is the shear modulus of the resin-rich layer. 

For pure Mode II loading, i.e., the End-Notched Flexure (ENF) model, the top and bottom arms are also subjected to different boundary conditions. Consequently, the contributions of the two arms must be considered separately in the pure Mode I case. The total sliding displacement of the cohesive element can therefore be written as:
\begin{align}\label{eq:DeltaI_total-mode-II}
    \Delta^{\textrm{total}}_{\textrm{II}} &= \sum_{i=1}^{n_{\textrm{top}}}\, \varepsilon_{xz}(z_i)\, h_{\textrm{rr}} +  \sum_{j=1}^{n_{\textrm{bot}}}\, \varepsilon_{xz}(z_j)\, h_{\textrm{rr}} \nonumber\\[0.5em]
    &= \sum_{i=1}^{n_{\textrm{top}}}\, \frac{\tau_{xz}(z_i)}{G_{\textrm{rr}}}\, h_{\textrm{rr}} +  \sum_{j=1}^{n_{\textrm{bot}}}\, \frac{\tau_{xz}(z_j)}{G_{\textrm{rr}}}\, h_{\textrm{rr}}
\end{align}

According to the Mode II traction-separation law, the maximum transverse shear stress can be derived as follows:
\begin{align}\label{eq:tau_max}
   \tau_{xz}^{\textrm{max}} = K_{s} \, \Delta^{\textrm{total}}_{\textrm{II}} 
\end{align}
where $\tau_{xz}^{\textrm{max}}$ is the maximum transverse shear stress, and $K_{s}$ is the shear penalty stiffness.

Substituting Equations~\ref{eq:shear_stress-shear_strain} and \ref{eq:DeltaI_total-mode-II} into Equation~\ref{eq:tau_max}, the expression for the shear penalty stiffness is given by:
\begin{align}\label{eq:Ks_final}
K_{s} &= \frac{\tau_{xz}^{\textrm{max}}}
{\displaystyle\sum_{i=1}^{n_{\textrm{top}}}\, \frac{\tau_{xz}(z_i)}{G_{\textrm{rr}}}\, h_{\textrm{rr}} +  \sum_{j=1}^{n_{\textrm{bot}}}\, \frac{\tau_{xz}(z_j)}{G_{\textrm{rr}}}\, h_{\textrm{rr}}} \nonumber\\[0.5em]
& = \frac{1}
{\displaystyle\sum_{i=1}^{n_{\textrm{top}}}\, \frac{\tau_{xz}(z_i)}{\tau_{xz}^{\textrm{max}}}\, \frac{h_{\textrm{rr}}}{G_{\textrm{rr}}} +  \sum_{j=1}^{n_{\textrm{bot}}}\, \frac{\tau_{xz}(z_j)}{\tau_{xz}^{\textrm{max}}}\,\frac{h_{\textrm{rr}}}{G_{\textrm{rr}}}}
\end{align}

Since $h_{\textrm{rr}}/G_{\textrm{rr}}$ is a known constant for a given resin-rich layer, the determination of the shear penalty stiffness depends on the through-thickness shear stress ratios, $\tau_{xz}(z_i)/\tau_z^{\textrm{max}}$, in Equation~\ref{eq:Ks_final}.

In the symmetric ENF model, the top and bottom arms contain the same number of plies. Consequently, the transverse shear stress reaches its maximum value at the mid-plane and vanishes at the upper and lower free surfaces:
\begin{align}
    &\tau_{xz}(0) = \tau_{xz}^{\textrm{max}} \label{eq:BC-ENF-MID}\\[0.5em]
    &\tau_{xz}(\pm h) = 0  \label{eq:BC-ENF-TOP}
\end{align}

Although the top and bottom arms have the same laminate configuration, they are subjected to different loading conditions. Following the procedure adopted for the DCB model, the transverse shear stress distribution is derived separately for the top and bottom arms of the ENF model. Enforcing the boundary condition in Equation~\ref{eq:BC-ENF-MID} into the quadratic expression for the shear in Equation~\ref{eq:tau_xy-2nd} gives:
\begin{align}
    a_3 = \tau_{xz}^{\textrm{max}}
\end{align}

Similarly, the maximum shear stress is normalized to unity. Applying the free-surface boundary condition at the upper surface of the top arm and substituting coefficient $a_3$ into Equation~\ref{eq:tau_xy-2nd} yields the following relation between the remaining unknown coefficients:
\begin{align}\label{eq:coefficient-a1-a2-ENF-1}
    a_1h^2 + a_2h + 1 = 0
\end{align}

The remaining unknown coefficients, $a_1$ and $a_2$, are determined from the boundary conditions of the out-of-plane normal stress in the ENF model. Since the out-of-plane normal stress vanishes at both the mid-plane and the free surfaces, the out-of-plane normal stress satisfies:
\begin{align}
    & \sigma_{z}(0) = 0  \label{eq:BC-ENF-MID-normal}\\[0.5em]
    &\sigma_{z}(\pm h) = 0 \label{eq:BC-ENF-TOP-normal}
\end{align}

Applying the above boundary conditions to Equation~\ref{eq:sigma_y-3rd} yields the coefficient $a_4$ and an additional relationship between the remaining unknown coefficients $a_1$ and $a_2$:
\begin{align}
    &a_4 = 0 \label{eq:coefficient-a4-ENF} \\[0.5em]
    &\frac{a_1}{3}h^3+\frac{a_2}{2}h^2+h = 0\label{eq:coefficient-a1-a2-ENF-2}
\end{align}

Combining Equations~\ref{eq:coefficient-a1-a2-ENF-1} and \ref{eq:coefficient-a1-a2-ENF-2}, together with the requirement that the shear stress reaches its maximum value at the mid-plane, leads to a positive value of $a_1$. The coefficients $a_1$ and $a_2$ are given by:
\begin{align}
    a_1 &= \frac{3}{h^2} \label{eq:coefficient-a1-ENF} \\[0.5em]
    a_2 &= -\frac{4}{h} \label{eq:coefficient-a2-ENF}
\end{align}

Substituting these coefficients into Equations~\ref{eq:tau_xy-2nd} and \ref{eq:sigma_y-3rd} gives the following stress distributions in the top arm of the ENF model:
\begin{align}
    \tau_{xz}^{\textrm{top}} &\propto\frac{3}{h^2}z^2-\frac{4}{h}z + 1\label{eq:tau_xy-ENF-top}  \\[0.5em]
    \sigma_{z}^{\textrm{top}} &\propto \frac{1}{h^2}z^3-\frac{2}{h}z^2+z \label{eq:sigma_y-ENF-top}
\end{align}

Because the transverse shear stress distribution in the ENF model is symmetric about the mid-plane, the coefficients for the bottom arm can be obtained directly from those of the top arm. The coefficient $a_1$ remains unchanged, whereas $a_2$ has the same magnitude but the opposite sign. The coefficients $a_3$ and $a_4$, determined from the mid-plane boundary conditions, are identical for both arms. Accordingly, the coefficients for the bottom arm are:
\begin{align}
\text{ENF bottom arm:}\qquad\left\{
\begin{aligned}
   a_1 &= \frac{3}{h^2} \\[0.5em]  
   a_2 &= \frac{4}{h} \\[0.5em] 
   a_3 &= 1 \\[0.5em] 
   a_4 &= 0  
\end{aligned}
\right.
\end{align}

Accordingly, the corresponding stress distributions in the bottom arm can be expressed as:
\begin{align}
    \tau_{xz}^{\textrm{bot}} &\propto\frac{3}{h^2}z^2+\frac{4}{h}z +1\label{eq:tau_xy-ENF-bot}  \\[0.5em]
    \sigma_{z}^{\textrm{bot}} &\propto \frac{1}{h^2}z^3+\frac{2}{h}z^2+z \label{eq:sigma_y-ENF-bot}
\end{align}

Combining Equations~\ref{eq:tau_xy-ENF-top} and \ref{eq:tau_xy-ENF-bot} gives the following piecewise expression for the transverse shear stress distribution under pure Mode II loading:
\begin{align}\label{eq:tau_xy-ENF-Final}
    \tau_{xz}^{\textrm{Mode II}} \propto \begin{cases}
   \dfrac{3}{h^2}z^2-\dfrac{4}{h}z + 1 , & h\ge z \ge 0 \\[1em]
   \dfrac{3}{h^2}z^2+\dfrac{4}{h}z +1  , & 0> z \ge-h
\end{cases}
\end{align}

Similarly, combining Equations~\ref{eq:sigma_y-ENF-top} and \ref{eq:sigma_y-ENF-bot} yields the following expression for the out-of-plane normal stress distribution:
\begin{align}\label{eq:sigma_y-ENF-Final}
    \sigma_{z}^{\textrm{Mode II}} \propto \begin{cases}
    \dfrac{1}{h^2}z^3-\dfrac{2}{h}z^2+z, & h\ge z \ge 0 \\[1em]
    \dfrac{1}{h^2}z^3+\dfrac{2}{h}z^2+z   , & 0> z \ge-h
\end{cases}
\end{align}

\begin{figure}[h!]
	\centering
	\includegraphics[width=1\linewidth]{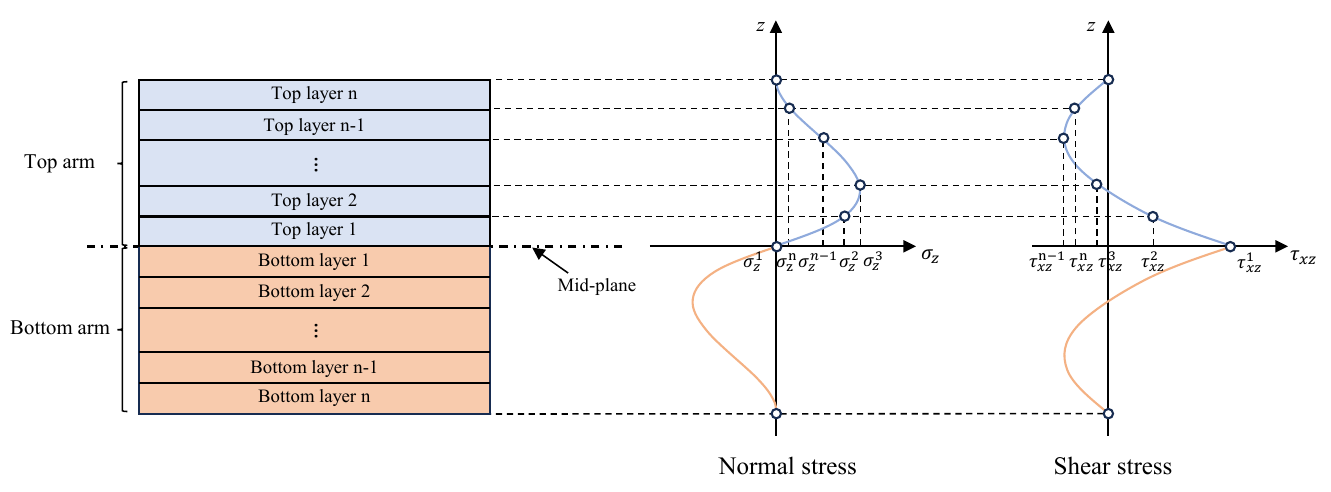}
	\caption{Stress distributions in the laminate under pure Mode II loading}
	\label{fig:SigmaY-TauXY-distribution-ENF}
\end{figure}

The through-thickness distributions of the transverse shear stress and out-of-plane normal stress under pure Mode II loading, as described by Equations~\ref{eq:tau_xy-ENF-Final} and \ref{eq:sigma_y-ENF-Final} are shown in Figure~\ref{fig:SigmaY-TauXY-distribution-ENF}. In the figure, $\sigma_z^i$ and $\tau_{xz}^i$ are the values of $\sigma_z^{\textrm{Mode II}}$ and $\tau_{xz}^{\textrm{Mode II}}$, respectively, at the $i^{\textrm{th}}$ interface layer.

The expression for the transverse shear stress directly provides the stress ratios $\tau_{xz}(z_i)/\tau_{xz}^{\textrm{max}}$ shown in Equation~\ref{eq:Ks_final}. Substituting these ratios into Equation~\ref{eq:Ks_final} yields the corresponding shear penalty stiffness.

\section{Validation}\label{sec:validation}
The proposed penalty stiffness formulation is first validated against three classical delamination benchmarks: the Double Cantilever Beam (DCB), End-Notched Flexure (ENF), and Mixed-Mode Bending (MMB). The numerical predictions are systematically compared with analytical solutions~\cite{williams1989fracture,ASTMD5528,ASTMD6671/D6671M} and finite element results using the conventional penalty stiffness. In addition, to highlight the advantages of the proposed method in compressive stress prediction, the traction distribution at the crack tip in the DCB model is examined and compared with the results reported in Reference~\cite{balducci2024overcoming}.

In addition to the classical delamination benchmarks, the Reinforced Double Cantilever Beam (R-DCB) model is considered to further validate the proposed penalty stiffness for fully three-dimensional delamination problems. The predictions are compared with experimental data and numerical results obtained using the penalty stiffness proposed in the previous study~\cite{ai2025structural}. This section first introduces the problem descriptions and modeling details, followed by comparisons of load-displacement curves and damage maps.

\subsection{Mode I delamination benchmark: Double Cantilever Beam (DCB)}\label{subsec:DCB}
\subsubsection{Description of the DCB specimen}\label{subsubsec:DCB-description}
\begin{figure}[h!]
	\centering
	\includegraphics[width=0.85\linewidth]{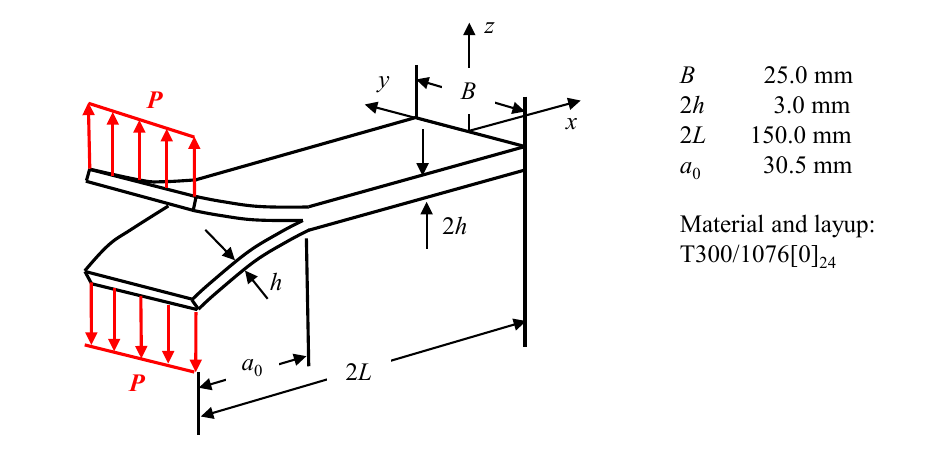}
	\caption{Double cantilever beam (DCB) specimen~\cite{krueger2015summary}}
	\label{fig:DCB-model}
\end{figure}

The DCB specimen, illustrated in Figure~\ref{fig:DCB-model}, is selected as the benchmark for
Mode I delamination analysis because it exhibits pure Mode I fracture, with $(G_{\text{II}}/G_{\text{T}})=0$. The geometric parameters and boundary conditions are presented in Figure~\ref{fig:DCB-model}. The laminate is manufactured from T300/1076 composite material, and the corresponding material properties are summarized in Table~\ref{tab:T300Material}.
\begin{table}[h!]
  \centering
  \caption{Material properties for the DCB specimen~\cite{krueger2015summary}}
    \begin{tabular}{lll}
    \toprule
 \multicolumn{3}{l}{\textbf{T300/1076 Unidirectional graphite/epoxy prepreg}}  \\
 \midrule
    $E_{11}$= 139.4 GPa & $E_{22}$= 10.16 GPa & $E_{33}$ = 10.16 GPa \\[0.3em]
    $\nu_{12}$ = 0.30 & $\nu_{13}$ = 0.30 & $\nu_{23}$ = 0.436  \\[0.3em]
    $G_{12}$ = 4.6 GPa & $G_{13}$ = 4.6 GPa & $G_{23}$ = 3.54 GPa \\[0.5em]
    Fracture toughness data & & \\[0.3em]
    $G_{\rm{Ic}} = 0.170 \, \mathrm{kJ/m^2} $ & $G_{\rm{IIc}} = 0.494 \, \mathrm{kJ/m^2}$ & $\eta = 1.62$ \\ [0.3em]
    Interfacial strength data \cite{turon2010accurate,lu2019cohesive} & & \\[0.3em]
    $\tau_{\rm{Ic}} = 30 \, \mathrm{MPa} $ & $\tau_{\rm{IIc}} = 60 \, \mathrm{MPa}$ &  \\ 
    \bottomrule
    \end{tabular}%
  \label{tab:T300Material}%
\end{table}%

\begin{figure}[h!]
	\centering
	\includegraphics[width=0.95\linewidth]{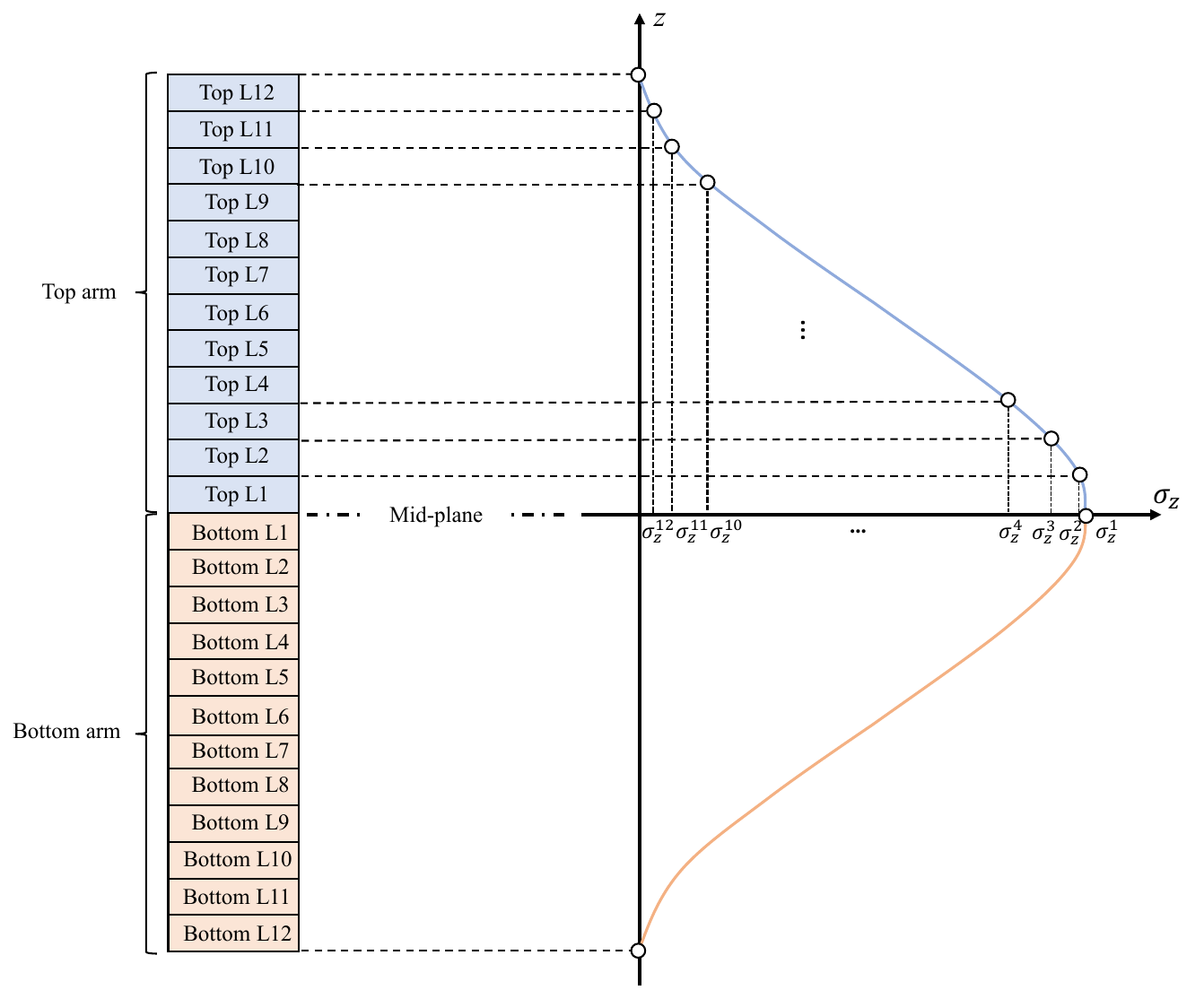}
	\caption{Through-thickness distribution of out-of-plane normal stress in the 24-ply DCB specimen}
	\label{fig:Stress-12Layers}
\end{figure}

All 24 plies in this benchmark have a fiber orientation of $0^{\circ}$. Consequently, the laminate can be represented using one shell element layer for each arm and a single layer of structural cohesive elements located at the delamination interface. According to Equation~\ref{eq:Kn_final}, the normal penalty stiffness depends on the number of laminate plies. Since both the top and bottom arms contain 12 plies, the corresponding out-of-plane normal stress distribution through the thickness for the 24-ply DCB specimen is shown in Figure~\ref{fig:Stress-12Layers}.

From the out-of-plane normal stress distribution in Figure~\ref{fig:Stress-12Layers} and Equation~\ref{eq:sigma_y-DCB-Final}, the stress ratios $\sigma_z(z_i)/\sigma_z^{\textrm{max}}$ for the interface layers can be readily evaluated. The corresponding values for the top arm are tabulated in Table~\ref{tab:Hrr-12Layer-normal}. 
\begin{table}[h!]
  \centering
  \caption{Out-of-plane normal stress ratios at the interface layers of the top arm in the 24-ply DCB specimen}
    \begin{tabular}{ccccccccccccc}
    \toprule
 $\dfrac{\sigma_z^1}{\sigma_z^{\textrm{max}}}$  & $\dfrac{\sigma_z^2}{\sigma_z^{\textrm{max}}}$ & $\dfrac{\sigma_z^3}{\sigma_z^{\textrm{max}}}$ &  $\dfrac{\sigma_z^4}{\sigma_z^{\textrm{max}}}$ &
 $\dfrac{\sigma_z^5}{\sigma_z^{\textrm{max}}}$&
 $\dfrac{\sigma_z^{\textrm{6}}}{\sigma_z^{\textrm{max}}}$ & $\dfrac{\sigma_z^7}{\sigma_z^{\textrm{max}}}$
  & $\dfrac{\sigma_z^8}{\sigma_z^{\textrm{max}}}$ & $\dfrac{\sigma_z^9}{\sigma_z^{\textrm{max}}}$ & $\dfrac{\sigma_z^{10}}{\sigma_z^{\textrm{max}}}$ & $\dfrac{\sigma_z^{11}}{\sigma_z^{\textrm{max}}}$ & $\dfrac{\sigma_z^{12}}{\sigma_z^{\textrm{max}}}$
    \\[0.8em]
 \midrule
   1.0 & 0.98 & 0.93 & 0.84 & 0.74 & 0.62 & 0.5 & 0.38 & 0.26 & 0.16 & 0.07 & 0.02
 \\[0.2em]
    \bottomrule
    \end{tabular}%
  \label{tab:Hrr-12Layer-normal}%
\end{table}

Therefore, the sum of the stress ratios for the top arm is given by:
\begin{align}\label{eq:total-hrr-normal}
\sum_{i=1}^{n_{\textrm{top}}^{\textrm{DCB}}}\, \frac{\sigma_z(z_i)}{\sigma_z^{\textrm{max}}} = \frac{\sigma_z^1+\sigma_z^2+\cdots+\sigma_z^{12}}{\sigma_z^{\textrm{max}}} = 6.5
\end{align}
where $n_{\textrm{top}}^{\textrm{DCB}}$ is the number of plies in the top arm of the DCB specimen.

Because the 24-ply DCB specimen is symmetric with respect to the mid-plane, the total sum of stress ratios in Equation~\ref{eq:Kn_final} becomes:
\begin{align}
    \sum_{i=1}^{n_{\textrm{top}}^{\textrm{DCB}}}\, \frac{\sigma_z(z_i)}{\sigma_z^{\textrm{max}}} +  \sum_{j=1}^{n_{\textrm{bot}}^{\textrm{DCB}}}\, \frac{\sigma_z(z_j)}{\sigma_z^{\textrm{max}}} &=   2\sum_{i=1}^{12}\, \frac{\sigma_z(z_i)}{\sigma_z^{\textrm{max}}}  \nonumber\\[0.5em]
    &= 2\times6.5 =13
\end{align}
where $n_{\textrm{bot}}^{\textrm{DCB}}$ is the number of plies in the bottom arm of the DCB specimen.

Because the elastic modulus $E^{\textrm{DCB}}_{\textrm{rr}}$ and Poisson's ratio $\nu^{\textrm{DCB}}_\textrm{rr}$ of the 1076 resin-rich layer are unavailable in the literature, approximate values are adopted from Crews et al.~\cite{crews1988fiber}. The DCB specimen considered in that study consisted of a 24-ply unidirectional graphite/epoxy laminate with a geometry similar to that investigated in this work. Accordingly, the reported material properties are expected to provide a reasonable approximation for the present DCB model and are listed in Table~\ref{tab:Kvalue-DCB}. The resin-rich layer thickness $h_{\textrm{rr}}$ is obtained from Grande et al.~\cite{grande1991effects} and the corresponding value is also given in Table~\ref{tab:Kvalue-DCB}.

\begin{table}[h!]
  \centering
  \caption{Material parameters of the resin-rich layer in the DCB specimen~\cite{crews1988fiber,grande1991effects}}
    \begin{tabular}{cccc}
    \toprule
    $E^{\textrm{DCB}}_{\textrm{rr}}$ $\mathrm{(N/mm^2)}$ & $G^{\textrm{DCB}}_{\textrm{rr}}$ $\mathrm{(N/mm^2)}$ & $\nu^{\textrm{DCB}}_\textrm{rr}$  & $h_{\textrm{rr}}$ $\mathrm{(mm)}$\\[0.3em]
 \midrule
    3400 & 1300 & 0.3 & 0.02286 \\[0.3em]
    \bottomrule
    \end{tabular}%
  \label{tab:Kvalue-DCB}%
\end{table}

Based on Equation~\ref{eq:Kn_final}, the normal penalty stiffness of the DCB specimen can be calculated as follows:
\begin{align}\label{eq:Kn_DCB}
K^{\textrm{DCB}}_{n}  &= \frac{1}
{\displaystyle\sum_{i=1}^{12}\, \frac{\sigma_z(z_i)}{\sigma_z^{\textrm{max}}}\, \frac{h_{\textrm{rr}}}{E^{\textrm{DCB}}_{\textrm{rr}}} +  \sum_{j=1}^{12}\, \frac{\sigma_z(z_j)}{\sigma_z^{\textrm{max}}}\,\frac{h_{\textrm{rr}}}{E^{\textrm{DCB}}_{\textrm{rr}}}} \nonumber\\[0.5em]
& = \frac{1}{13\times\frac{0.02286}{3400}} = 11441 \, \mathrm{N/mm^3} 
\end{align}

Because the DCB specimen represents a pure Mode I problem, the shear penalty stiffness has a negligible effect on the numerical predictions. Nearly identical results are obtained for different values of this parameter. Accordingly, the shear penalty stiffness is not considered further in this section.

\subsubsection{Load-displacement curves}\label{subsubsec:RF-U-DCB}
\begin{figure}[h!]
      \centering
	   \begin{subfigure}{0.48\linewidth}
		\includegraphics[scale=1]{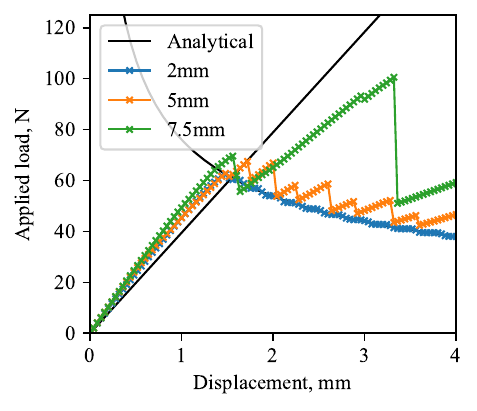}
		\caption{Conventional penalty stiffness}
		\label{fig:DCB-comparison-Turon}
	   \end{subfigure}
	   \begin{subfigure}{0.48\linewidth}
		\includegraphics[scale=1]{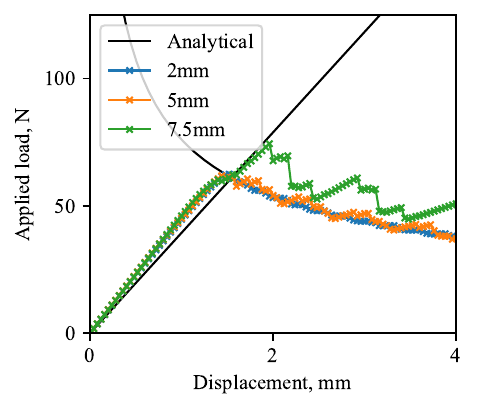}
		\caption{Proposed penalty stiffness}
		\label{fig:DCB-comparison-XAI}
	    \end{subfigure}
	\caption{Comparison of the load-displacement curves obtained from the DCB simulations using the conventional and proposed penalty stiffness formulations at different mesh densities}
	\label{fig:DCB-comparison}
\end{figure}

The load-displacement curves obtained from the numerical simulations are shown in Figure~\ref{fig:DCB-comparison}. The left panel corresponds to the conventional penalty stiffness, while the right panel corresponds to the proposed penalty stiffness. In both cases, the numerical results are compared with the analytical solution, represented by the black curve. 

As shown in Figure~\ref{fig:DCB-comparison-Turon}, the model using the conventional penalty stiffness fails to achieve mesh convergence when the element size exceeds 5 mm. In contrast, the proposed penalty stiffness maintains good convergence even with a 7.5 mm mesh size (Figure~\ref{fig:DCB-comparison-XAI}). Comparing the results obtained with the 5 mm mesh in the two figures, the proposed normal penalty stiffness produces a load-displacement curve with smaller oscillations and better agreement with the analytical solution than the conventional penalty stiffness. It demonstrates that the proposed formulation significantly reduces the numerical error associated with coarser meshes. For finer meshes, such as the 2 mm mesh, both penalty stiffness formulations produce results that are in close agreement with the analytical solution, with no significant differences observed. In summary, the proposed normal penalty stiffness significantly improves mesh convergence and allows accurate predictions with coarser meshes, thereby enhancing computational efficiency.

\subsubsection{Normalized $L_2$ error analysis}\label{subsubsec:Error-analysis}
To quantitatively evaluate the effects of the penalty stiffness and mesh convergence, the normalized $L_2$ error is adopted. It measures the discrepancy between the numerical load-displacement curve, denoted as $f^*(x)$, and the corresponding analytical solution $f(x)$. The normalized $L_2$ error is defined as follows:
\begin{align}\label{eq:L2 error}
    e=\frac{\sqrt{\int_a^b(f^*-f)^2\dd x}}{\sqrt{\int_a^bf^2\dd x}}
\end{align}

\begin{figure}[h!]
	\centering
	\includegraphics[scale=1]{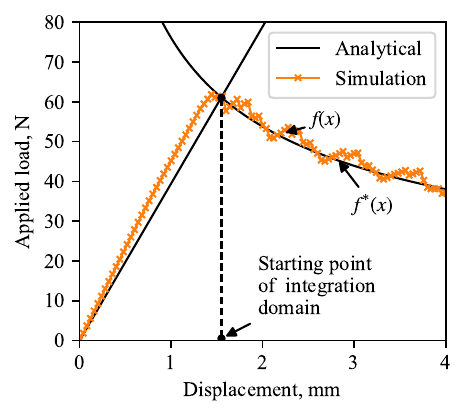}
	\caption{Definition of the integration interval for the normalized $L_2$ error calculation}
	\label{fig:L2-Error-diagram}
\end{figure}

Due to the fact that the numerical and analytical responses are in close agreement within the elastic region, and obvious differences arise only after the peak load, the lower integration limit $a$ in Equation~\ref{eq:L2 error} is set to the displacement corresponding to the peak load (indicated by the black dot in Figure~\ref{fig:L2-Error-diagram}). The upper integration limit $b$ is taken as the maximum displacement of the DCB simulation, i.e., 4 mm.

\begin{figure}[h]
	\centering
	\includegraphics[scale=1]{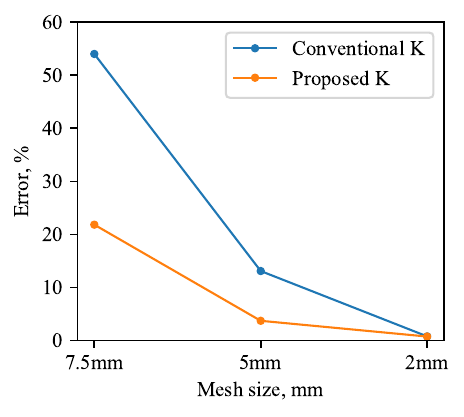}
	\caption{Normalized $L_2$ error of the DCB benchmark versus mesh size for different penalty stiffness formulations}
	\label{fig:L2-Error-dcb}
\end{figure}

In this study, the trapezoidal rule is applied to numerically evaluate the integral in Equation~\ref{eq:L2 error}. The integration interval is discretized according to the displacement points of the numerical load-displacement curves, and the integrand is assumed to vary linearly within each subinterval. The normalized $L_2$ error is then calculated by summing the areas of the trapezoids. The calculated errors are presented in Figure~\ref{fig:L2-Error-dcb}. From the figure, when the conventional penalty stiffness is adopted, the normalized $L_2$ error exceeds 50\% for the 7.5 mm mesh, which is markedly larger than that obtained using the proposed normal penalty stiffness. As the mesh is progressively refined to 2 mm, the influence of the penalty stiffness reduces, and the errors predicted by the two formulations are approximately 0.8\%.

\subsubsection{Normal traction distribution of DCB crack tip}\label{subsubsec:Traction-DCB}
\begin{figure}[h]
	\centering
	\includegraphics[width=0.80\linewidth]{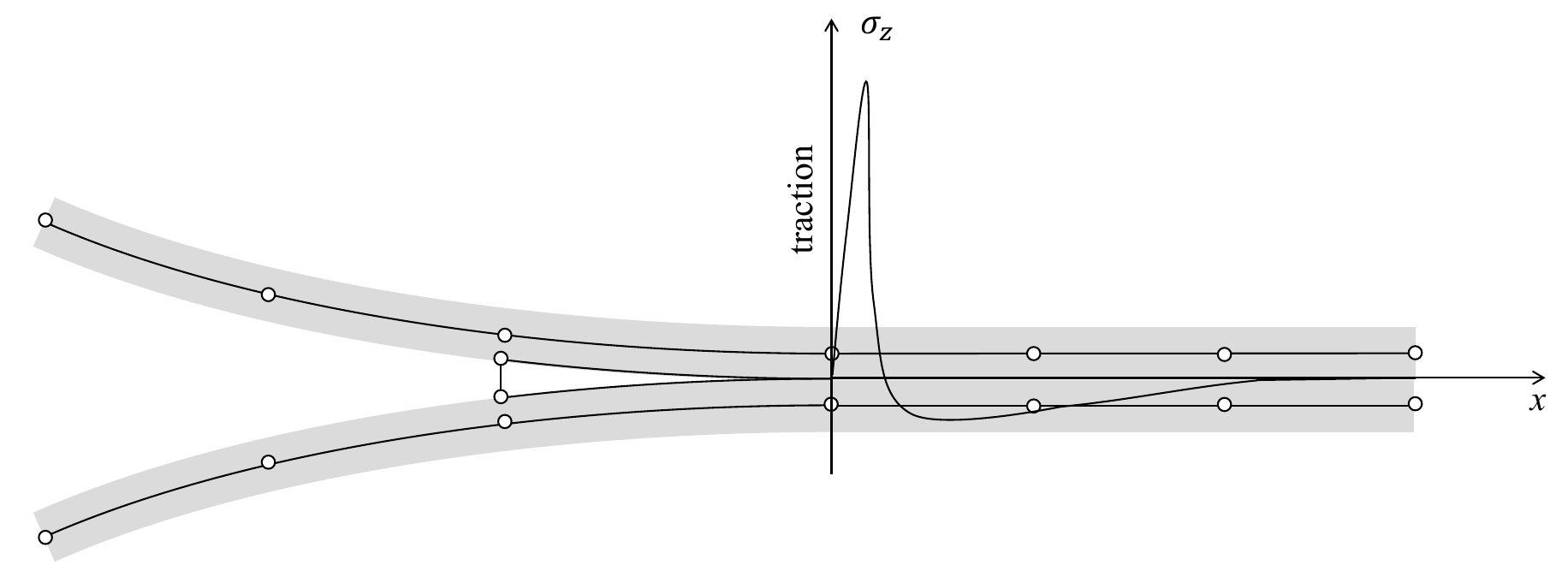}
	\caption{Distribution of traction ahead of crack tip}
	\label{fig:General-DCB-Traction}
\end{figure}

Figure~\ref{fig:General-DCB-Traction} presents the traction distribution ahead of the crack tip at the onset of delamination in the DCB specimen. As predicted by cohesive zone theory, the traction at the crack tip reduces to zero, where damage initiates. Subsequently, the traction increases in the direction of crack propagation until it reaches its maximum value. The distance between the crack tip and the location of the maximum traction is defined as the cohesive zone length (CZL). Beyond the maximum traction point, the traction decreases and becomes compressive. Then, the traction reaches a minimum value before eventually returning to zero.

In this section, the traction distribution in the DCB specimen is evaluated. The predicted results are compared with those obtained from the higher-order plate (TUBA3) model and a solid element model under a very refined mesh, as reported by Tosti Balducci et al.~\cite{balducci2024overcoming}.
The traction extraction procedure and the corresponding comparison results are presented below.
\begin{figure}[h]
	\centering
	\includegraphics[width=0.95\linewidth]{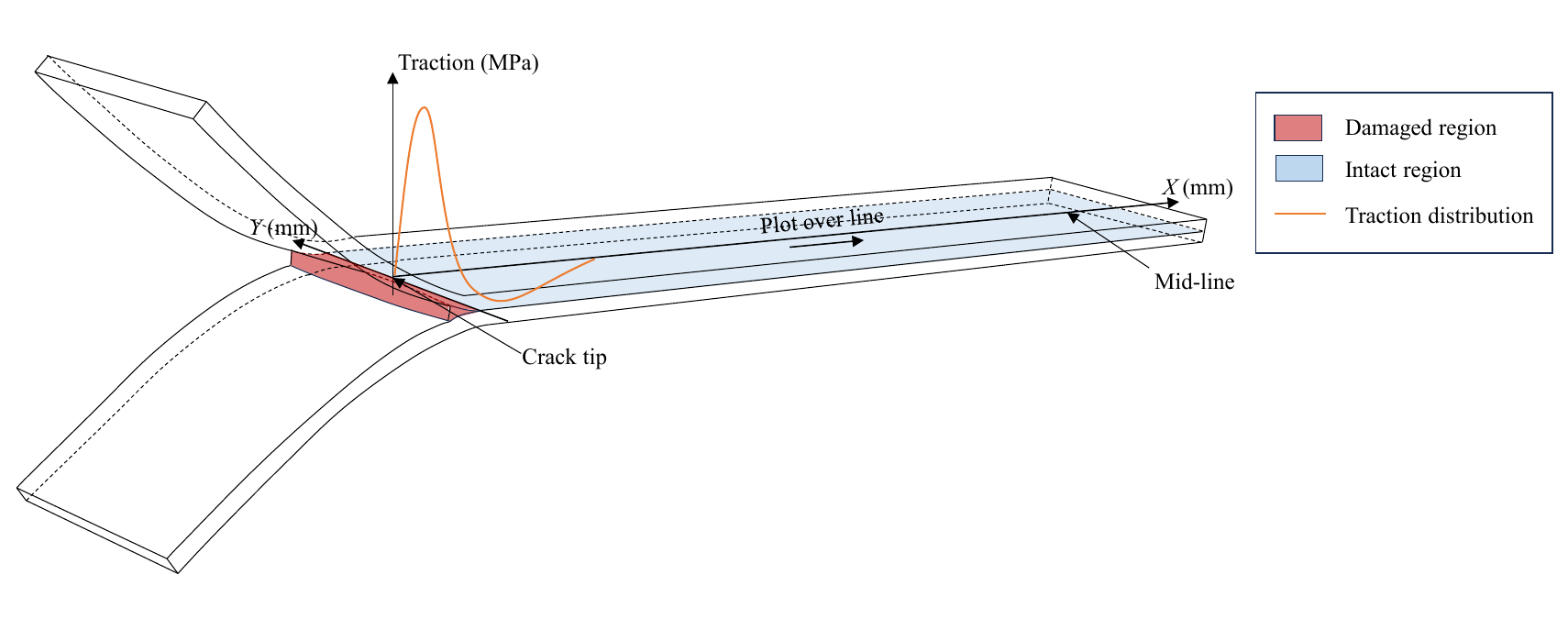}
	\caption{Procedure for extracting the traction distribution from cohesive elements}
	\label{fig:DCB-Traction-Interpolation}
\end{figure}

First, the traction values are extracted from the integration points of the cohesive elements (CEs) and stored for post-processing. Then, these values are imported into ParaView and visualized according to their global coordinates, denoted as $X$ and $Y$ in Figure~\ref{fig:DCB-Traction-Interpolation}. The traction distribution is subsequently evaluated by interpolating along the specimen mid-line in the $X\textrm{-}Y$ plane using the \textit{Plot Over Line} filter. This procedure is shown in Figure~\ref{fig:DCB-Traction-Interpolation} and enables a consistent evaluation of the traction distribution for finite element models with different mesh densities while minimizing the influence of free-edge effects.

\begin{figure}[h!]
      \centering
	   \begin{subfigure}{0.48\linewidth}
		\includegraphics[scale=1]{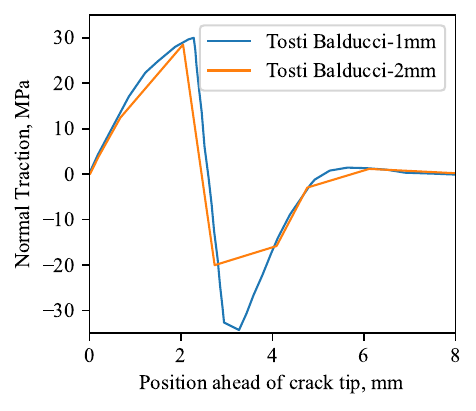}
		\caption{Conventional penalty stiffness}
		\label{fig:DCB-Traction-Giorgio}
	   \end{subfigure}
	   \begin{subfigure}{0.48\linewidth}
		\includegraphics[scale=1]{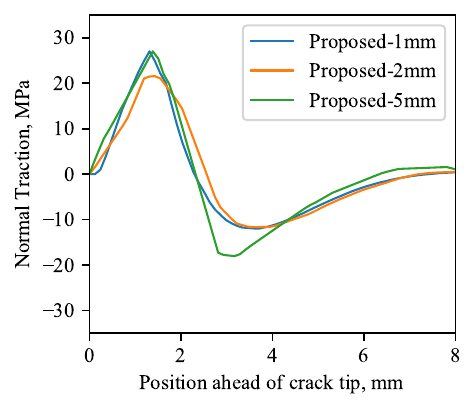}
		\caption{Proposed penalty stiffness}
		\label{fig:DCB-Traction-XAI}
	    \end{subfigure}
	\caption{Comparison of normal traction distributions in the cohesive zone of the DCB specimen for different mesh sizes using the conventional and proposed penalty stiffness formulations} 
	\label{fig:DCB-Comparison-XAI2Giorgio}
\end{figure}

Figure~\ref{fig:DCB-Comparison-XAI2Giorgio} compares the normal traction distributions ahead of the crack tip predicted by the DCB simulations. The left panel shows the results from the TUBA3 element with the conventional penalty stiffness at different mesh sizes, while the right panel shows the corresponding results from the cubic structural cohesive element with the proposed normal penalty stiffness.

As shown in Figure~\ref{fig:DCB-Comparison-XAI2Giorgio}, both penalty stiffness formulations accurately capture the maximum traction, which is attributed to the identical peak traction of 30 MPa prescribed by the damage initiation criterion. A clear difference in mesh convergence is observed with respect to compression. For the conventional penalty stiffness, the predicted compression shows poor mesh convergence. Significant differences are observed between the 1 mm and 2 mm meshes, while the 5 mm mesh fails to converge and cannot reproduce the traction profile accurately. A similar lack of mesh convergence has also been reported for the two-dimensional DCB model based on higher-order beam elements~\cite{russo2020overcoming}. In contrast, the proposed penalty stiffness exhibits excellent mesh convergence. The predicted traction distributions show significantly more stable predictions as the mesh is refined from 5 mm to 1 mm, demonstrating that accurate traction predictions can be achieved even with coarse meshes.
\begin{figure}[h!]
	\centering
	\includegraphics[scale=1]{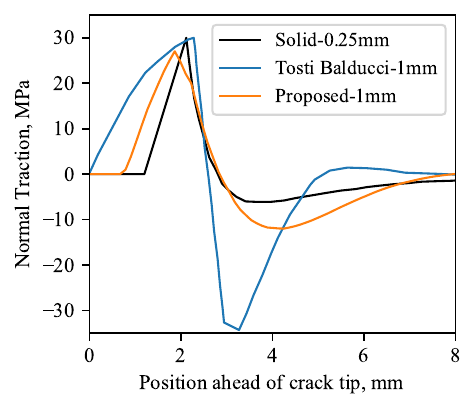}
	\caption{Comparison of the predicted normal traction distributions ahead of the crack tip}
	\label{fig:DCB-Comparison-1mm}
\end{figure}

Figure~\ref{fig:DCB-Comparison-1mm} compares the traction distributions predicted by three numerical models. The black curve represents the 3D Abaqus cohesive contact model with a mesh size of 0.25 mm. The blue and orange curves correspond to the 1 mm TUBA3 model with the conventional penalty stiffness and the structural cohesive model with the proposed penalty stiffness, respectively. The proposed penalty stiffness significantly reduces the predicted compression. The resulting compressive magnitude is approximately one-third of that predicted by the TUBA3 model. Nevertheless, it is still approximately two times greater than that predicted by the solid element model. This discrepancy is mainly attributed to the absence of transverse shear deformation in the structural ply element formulation. Further reducing the discrepancy with solid element models is beyond the scope of the present work and will be investigated in future research.

\subsection{Mode II delamination benchmark: End-Notched Flexure (ENF)}\label{subsec:ENF}
\subsubsection{Description of the ENF specimen}\label{subsubsec:ENF-description}
\begin{figure}[h!]
	\centering
	\includegraphics[width=0.85\linewidth]{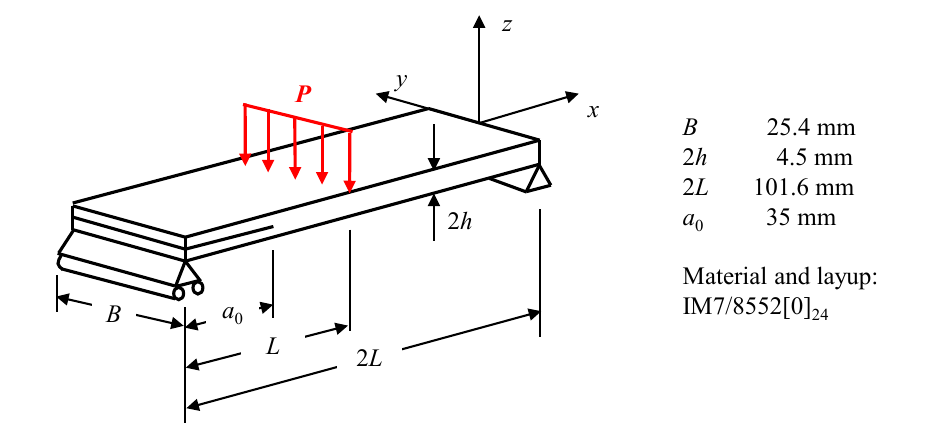}
	\caption{End-notched flexure (ENF) specimen~\cite{krueger2015summary}}
	\label{fig:ENF-model}
\end{figure}

The ENF specimen, shown in Figure~\ref{fig:ENF-model}, is selected as the Mode II delamination benchmark because it exhibits pure Mode II fracture ($G_{\text{II}}/G_{\text{T}} = 1$). The specimen geometry and boundary conditions are also illustrated in Figure~\ref{fig:ENF-model}. Unlike the DCB benchmark, the ENF specimen is fabricated from IM7/8552 composite laminate, whose material properties are listed in Table~\ref{tab:ENFmaterial}. The resin-rich layer is modelled using the 8552 epoxy material, whose material properties are taken from Krueger~\cite{krueger2006analysis}. The corresponding parameters are summarized in Table~\ref{tab:Kvalue-ENF}.
\begin{table}[h!]
  \centering
  \caption{Material properties of the ENF and MMB specimens~\cite{krueger2015summary}}
    \begin{tabular}{lll}
    \toprule
 \multicolumn{3}{l}{\textbf{IM7/8552 Unidirectional graphite/epoxy prepreg}}  \\
 \midrule
    $E_{11}$= 161 GPa & $E_{22}$= 11.38 GPa & $E_{33}$ = 11.38 GPa \\[0.3em]
    $\nu_{12}$ = 0.32 & $\nu_{13}$ = 0.32 & $\nu_{23}$ = 0.45  \\[0.3em]
    $G_{12}$ = 5.2 GPa & $G_{13}$ = 5.2 GPa & $G_{23}$ = 3.9 GPa \\[0.5em]
    Fracture toughness data & & \\[0.3em]
    $G_{\rm{Ic}} = 0.212 \, \mathrm{kJ/m^2} $ & $G_{\rm{IIc}} = 0.774 \, \mathrm{kJ/m^2}$ & $\eta = 2.1$ \\ [0.3em]
    Interfacial strength data \cite{turon2010accurate,lu2019cohesive} & & \\[0.3em]
    $\tau_{\rm{Ic}} = 30 \, \mathrm{MPa} $ & $\tau_{\rm{IIc}} = 60 \, \mathrm{MPa}$ &  \\ 
    \bottomrule
    \end{tabular}%
  \label{tab:ENFmaterial}%
\end{table}%


\begin{table}[h!]
  \centering
  \caption{Material parameters of 8552 epoxy resin~\cite{krueger2006analysis,grande1991effects}}
    \begin{tabular}{cccc}
    \toprule
    $E^{\textrm{ENF}}_{\textrm{rr}}$ $\mathrm{(N/mm^2)}$ & $G^{\textrm{ENF}}_{\textrm{rr}}$$\mathrm{(N/mm^2)}$  & $\nu^{\textrm{ENF}}_\textrm{rr}$&  $h_{\textrm{rr}}$ $\mathrm{(mm)}$  \\[0.3em]
 \midrule
    4700 & 1715 & 0.37 & 0.02286  \\[0.3em]
    \bottomrule
    \end{tabular}%
  \label{tab:Kvalue-ENF}%
\end{table}

\begin{figure}[h!]
	\centering
	\includegraphics[width=0.95\linewidth]{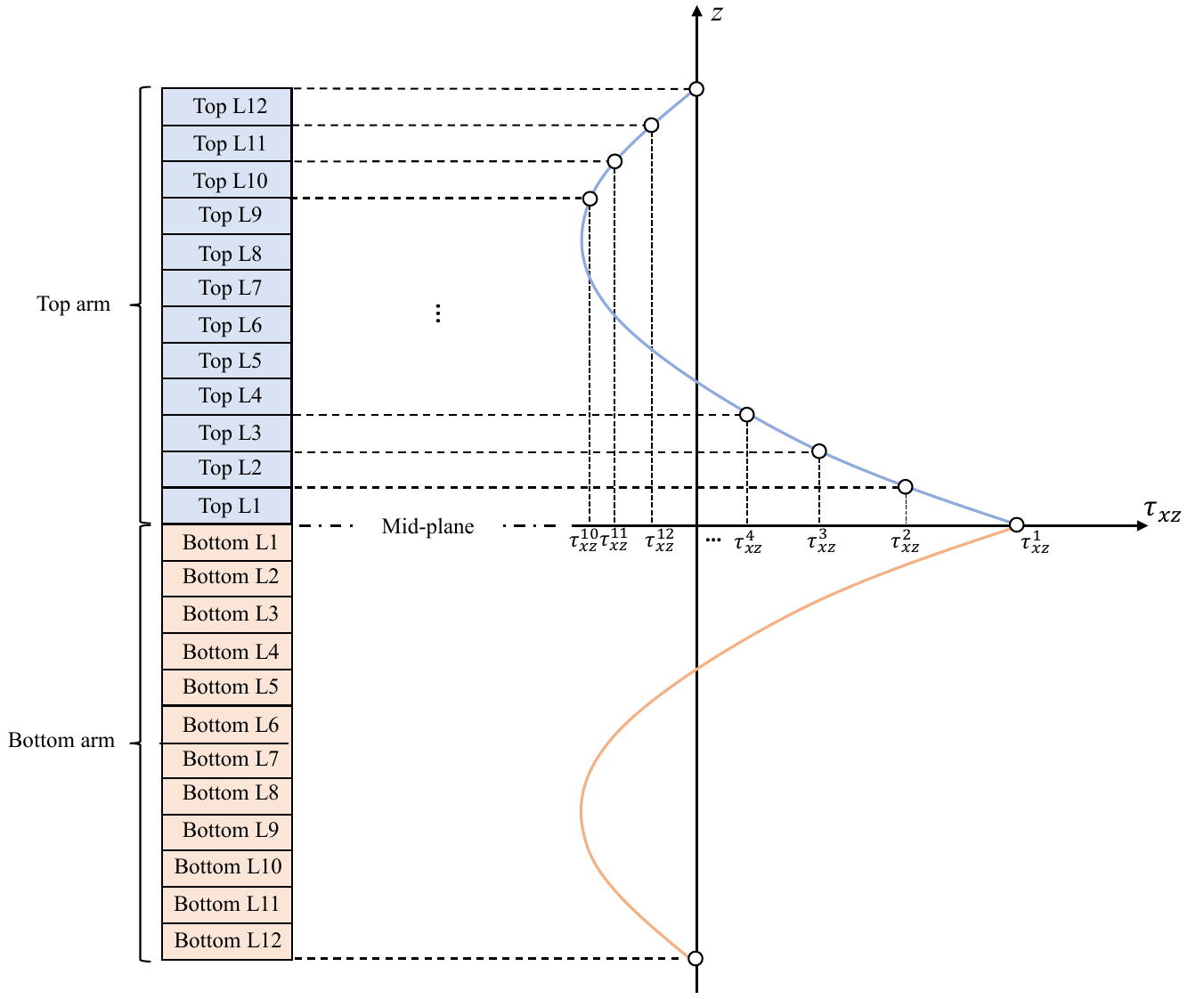}
	\caption{Through-thickness distribution of transverse shear stress in the 24-ply ENF specimen}
	\label{fig:Stress-12Layers-ENF}
\end{figure}

Similar to the DCB model, the ENF model consists of a 24-ply laminate with a $0^\circ$ layup throughout the specimen. According to Equation~\ref{eq:Ks_final}, the shear penalty stiffness is determined by taking into account the number of plies in each laminate arm. As the laminate is symmetric about the mid-plane, both the top and bottom arms contain 12 plies. The corresponding through-thickness transverse shear stress distribution is shown in Figure~\ref{fig:Stress-12Layers-ENF}.

Based on the transverse shear stress distribution shown in Figure~\ref{fig:Stress-12Layers-ENF} and Equation~\ref{eq:tau_xy-ENF-Final}, the stress ratio at each interface layer with respect to the maximum transverse shear stress is determined. The corresponding stress ratios for the interfaces in the top arm are summarized in Table~\ref{tab:Hrr-12Layer-shear}.
\begin{table}[h!]
  \centering
  \caption{Transverse shear stress ratios at the interface layers of the top arm in the 24-ply ENF specimen}
    \begin{tabular}{ccccccccccccc}
    \toprule
 $\dfrac{\tau_{xz}^1}{\tau_{xz}^{\textrm{max}}}$  & $\dfrac{\tau_{xz}^2}{\tau_{xz}^{\textrm{max}}}$ & $\dfrac{\tau_{xz}^3}{\tau_{xz}^{\textrm{max}}}$ &  $\dfrac{\tau_{xz}^4}{\tau_{xz}^{\textrm{max}}}$ &
 $\dfrac{\tau_{xz}^5}{\tau_{xz}^{\textrm{max}}}$&
 $\dfrac{\tau_{xz}^{\textrm{6}}}{\tau_{xz}^{\textrm{max}}}$ & $\dfrac{\tau_{xz}^7}{\tau_{xz}^{\textrm{max}}}$
  & $\dfrac{\tau_{xz}^8}{\tau_{xz}^{\textrm{max}}}$ & $\dfrac{\tau_{xz}^9}{\tau_{xz}^{\textrm{max}}}$ & $\dfrac{\tau_{xz}^{10}}{\tau_{xz}^{\textrm{max}}}$ & $\dfrac{\tau_{xz}^{11}}{\tau_{xz}^{\textrm{max}}}$ & $\dfrac{\tau_{xz}^{12}}{\tau_{xz}^{\textrm{max}}}$
    \\[0.8em]
 \midrule
   1.0 & 0.69 & 0.42 & 0.19 & 0.00 & -0.15 & -0.25 & -0.31 & -0.33 & -0.31 & -0.25 & -0.15
 \\[0.3em]
    \bottomrule
    \end{tabular}%
  \label{tab:Hrr-12Layer-shear}%
\end{table}

Therefore, the sum of the transverse shear stress ratios for the top arm is given by:
\begin{align}\label{eq:total-ENF-shear}
\sum_{i=1}^{n_{\textrm{top}}^{\textrm{ENF}}}\, \frac{\tau_{xz}(z_i)}{\tau_{xz}^{\textrm{max}}} = \frac{\tau_{xz}^1+\tau_{xz}^2+\cdots+\tau_{xz}^{12}}{\tau_{xz}^{\textrm{max}}} = 0.54
\end{align}
where $n_{\textrm{top}}^{\textrm{ENF}}$ is the number of plies in the top arm of the ENF specimen.

Because the 24-ply ENF specimen is symmetric about the laminate mid-plane, the sum of the stress ratios in Equation~\ref{eq:Ks_final} can be written as:
\begin{align}
    \sum_{i=1}^{n^{\textrm{ENF}}_{\textrm{top}}}\, \frac{\tau_{xz}(z_i)}{\tau_{xz}^{\textrm{max}}} +  \sum_{j=1}^{n^{\textrm{ENF}}_{\textrm{bot}}}\, \frac{\tau_{xz}(z_j)}{\tau_{xz}^{\textrm{max}}} &=   2\sum_{i=1}^{12}\, \frac{\tau_{xz}(z_i)}{\tau_{xz}^{\textrm{max}}} \nonumber\\[0.5em]
    &=  2\times 0.54=1.08
\end{align}
where $n_{\textrm{bot}}^{\textrm{ENF}}$ is the number of plies in the bottom arm of the ENF specimen.

According to Equation~\ref{eq:Ks_final}, the shear penalty stiffness of the ENF specimen can be calculated as:
\begin{align}\label{eq:Ks_ENF}
K^{\textrm{ENF}}_{s}  &= \frac{1}
{\displaystyle\sum_{i=1}^{12}\, \frac{\tau_{xz}(z_i)}{\tau_{xz}^{\textrm{max}}}\, \frac{h_{\textrm{rr}}}{G^{\textrm{ENF}}_{\textrm{rr}}} +  \sum_{j=1}^{12}\, \frac{\tau_{xz}(z_j)}{\tau_{xz}^{\textrm{max}}}\,\frac{h_{\textrm{rr}}}{G^{\textrm{ENF}}_{\textrm{rr}}}} \nonumber\\[0.5em]
& = \frac{1}{1.08\times\frac{0.02286}{1715}} = 69465 \, \mathrm{N/mm^3} 
\end{align}

\subsubsection{Load-displacement curves}\label{subsubsec:RF-U-ENF}
\begin{figure}[h!]
      \centering
	   \begin{subfigure}{0.48\linewidth}
		\includegraphics[scale=1]{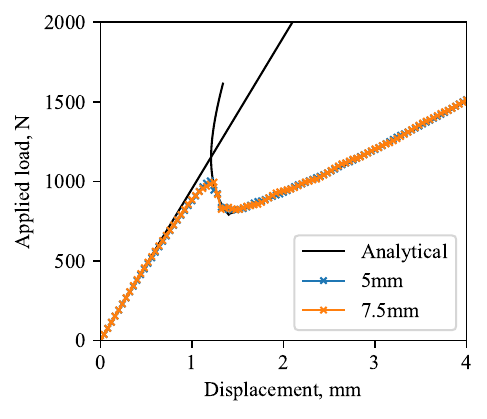}
		\caption{Conventional penalty stiffness}
		\label{fig:ENF-comparison-Turon}
	   \end{subfigure}
	   \begin{subfigure}{0.48\linewidth}
		\includegraphics[scale=1]{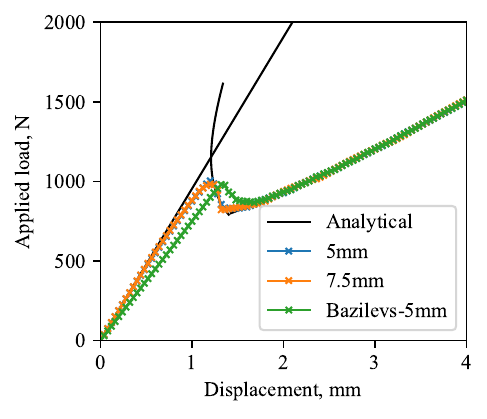}
		\caption{Proposed and Bazilevs' penalty stiffness}
		\label{fig:ENF-comparison-XAI}
	    \end{subfigure}
	\caption{Comparison of the load-displacement curves obtained from the ENF simulations using the conventional, Bazilevs', and proposed penalty stiffness formulations at different mesh densities}
	\label{fig:ENF-comparison}
\end{figure}

The load-displacement curves obtained from the ENF simulations are shown in Figure~\ref{fig:ENF-comparison}. The left panel presents the results obtained using the conventional penalty stiffness. The right panel compares the results obtained using the proposed shear penalty stiffness with those obtained using the formulation proposed by Bazilevs et al.~\cite{bazilevs2018new}. The latter is represented by the green curve, and the corresponding expression for the Bazilevs shear penalty stiffness is given below:
\begin{align}\label{eq:Bazilevs-K}
    K^{\textrm{Bazilevs}}_s &= \frac{G_{13}}{h_{\text{top}}/2 +h_{\text{bot}}/2} \nonumber\\[0.5em]
              &= \frac{5200}{2.25} = 2311 \, \mathrm{N/mm^3} 
\end{align}
where $G_{13}$ is the transverse shear modulus of the composite laminate.    

Figure~\ref{fig:ENF-comparison} compares the load--displacement curves obtained using the three shear penalty stiffness formulations. Both the conventional penalty stiffness and proposed shear penalty stiffness accurately capture the complete delamination process and exhibit excellent mesh convergence. Even with a coarse mesh size of 7.5 mm, the numerical predictions remain in close agreement with the analytical solution. In contrast, when Bazilevs's shear penalty stiffness is adopted, the initial stiffness is significantly underestimated. Although the predicted peak load is close to those obtained using the other two formulations, the corresponding displacement is noticeably larger than the analytical solution. This discrepancy is caused by the extremely low shear penalty stiffness defined in Equation~\ref{eq:Bazilevs-K}, which makes the structural response excessively compliant and reduces the accuracy of the predicted delamination process. A similar response was also reported by Bazilevs et al.~\cite{bazilevs2018new}. This result indicates that the improved performance of the present formulation is attributable to the proposed shear penalty stiffness rather than the shell element formulation

In summary, the proposed shear penalty stiffness does not provide a significant improvement over the conventional penalty stiffness. The difference between the proposed and the conventional formulations is less pronounced than in the DCB model. Nevertheless, the proposed formulation produces more accurate predictions than other shear penalty stiffness formulations developed for shell elements.

\subsection{Mixed-mode delamination benchmark: Mixed-Mode Bending (MMB)}\label{subsec:MMB}
\subsubsection{Description of the MMB specimen}\label{subsubsec:MMB-description}
\begin{figure}[h!]
	\centering
	\includegraphics[width=0.85\linewidth]{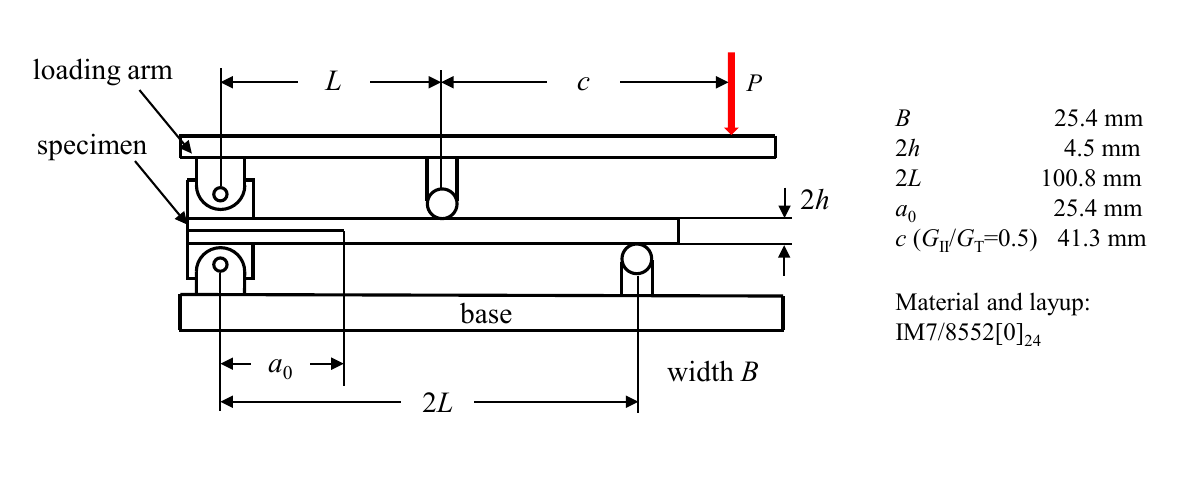}
	\caption{Mixed-mode bending (MMB) specimen~\cite{krueger2015summary}}
	\label{fig:MMB-model}
\end{figure}

The MMB specimen, shown in Figure~\ref{fig:MMB-model}, is designed for a mixed-mode ratio of $G_{\text{II}}/G_{\text{T}} = 0.5$. The laminate layup, specimen geometry, and initial crack length, $a_0$ are also illustrated in the figure. The material properties are identical to those of the ENF model and are listed in Table~\ref{tab:ENFmaterial}. Because the laminate configuration and material properties are identical to those of the ENF model, the shear penalty stiffness is also identical and is given by Equation~\ref{eq:Ks_final}:
\begin{align}
    K^{\textrm{MMB}}_s = K^{\textrm{ENF}}_s
    &= \frac{1}
{\displaystyle\sum_{i=1}^{12}\, \frac{\tau_{xz}(z_i)}{\tau_{xz}^{\textrm{max}}}\, \frac{h_{\textrm{rr}}}{G^{\textrm{ENF}}_{\textrm{rr}}} +  \sum_{j=1}^{12}\, \frac{\tau_{xz}(z_j)}{\tau_{xz}^{\textrm{max}}}\,\frac{h_{\textrm{rr}}}{G^{\textrm{ENF}}_{\textrm{rr}}}} \nonumber\\[0.5em]
& = \frac{1}{1.08\times\frac{0.02286}{1715}} = 69465 \, \mathrm{N/mm^3} 
\end{align}

For the normal penalty stiffness, the stress ratios in Equation~\ref{eq:Kn_final} are identical to those of the DCB model because the MMB and DCB laminates have the same configuration. The corresponding expression, calculated with the MMB material properties, is given by:
\begin{align}
K^{\textrm{MMB}}_{n}  &= \frac{1}
{\displaystyle\sum_{i=1}^{12}\, \frac{\sigma_z(z_i)}{\sigma_z^{\textrm{max}}}\, \frac{h_{\textrm{rr}}}{E^{\textrm{ENF}}_{\textrm{rr}}} +  \sum_{j=1}^{12}\, \frac{\sigma_z(z_j)}{\sigma_{\textrm{max}}}\,\frac{h_{\textrm{rr}}}{E^{\textrm{ENF}}_{\textrm{rr}}}} \nonumber\\[0.5em]
& = \frac{1}{13\times\frac{0.02286}{4700}} = 15815 \, \mathrm{N/mm^3} 
\end{align}

\subsubsection{Load-displacement curves}\label{subsubsec:RF-U-MMB}
\begin{figure}[h!]
      \centering
	   \begin{subfigure}{0.48\linewidth}
		\includegraphics[scale=1]{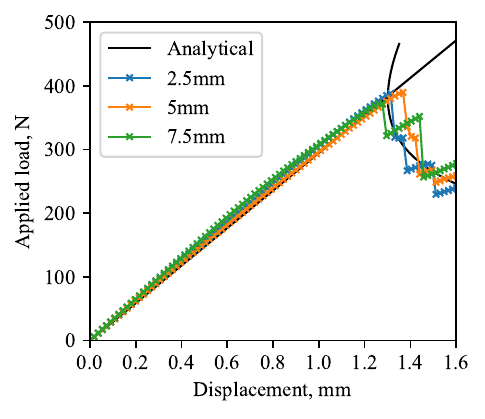}
		\caption{Conventional penalty stiffness}
		\label{fig:MMB-comparison-Turon}
	   \end{subfigure}
	   \begin{subfigure}{0.48\linewidth}
		\includegraphics[scale=1]{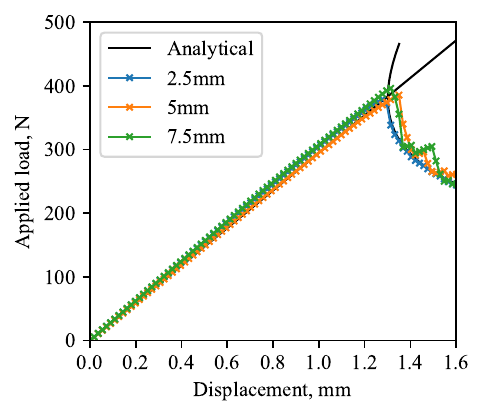}
		\caption{Proposed penalty stiffness}
		\label{fig:MMB-comparison-XAI}
	    \end{subfigure}
	\caption{Comparison of the load-displacement curves obtained from the MMB simulations using the conventional and proposed penalty stiffness formulations at different mesh densities}
	\label{fig:MMB-comparison}
\end{figure}

The load-displacement curves obtained using the proposed normal and shear penalty stiffness are shown in Figure~\ref{fig:MMB-comparison-XAI}. For comparison, the corresponding results obtained using the conventional penalty stiffness are presented in Figure~\ref{fig:MMB-comparison-Turon}.

A comparison of Figures~\ref{fig:MMB-comparison-XAI} and~\ref{fig:MMB-comparison-Turon} shows that the proposed penalty stiffness formulation produces smoother load-displacement curves than the conventional formulation. After the peak load, the predicted curves remain in excellent agreement with the analytical solution for all mesh sizes considered.  

\begin{figure}[h!]
	\centering
	\includegraphics[scale=1]{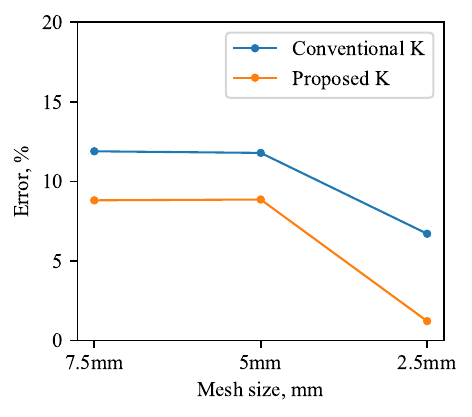}
	\caption{Normalized $L_2$ error of the MMB benchmark versus mesh size for different penalty stiffness formulations}
	\label{fig:L2-Error-MMB}
\end{figure}

To further quantify the improvement, the MMB specimen is evaluated using the same normalized $L_2$ error adopted for the DCB specimen. Figure~\ref{fig:L2-Error-MMB} compares the normalized $L_2$ error obtained using the conventional and proposed penalty stiffness formulations at different mesh sizes. As the mesh is refined from 7.5 mm to 2.5 mm, the error decreases for both formulations. Nevertheless, the proposed penalty stiffness consistently produces a lower normalized $L_2$ error than the conventional penalty stiffness. The improvement becomes more pronounced as the mesh is refined. In summary, the proposed penalty stiffness formulations reduce the prediction error and achieve closer agreement with the analytical solution for the MMB specimen.

\subsection{Three-dimensional delamination benchmark: Reinforced Double Cantilever Beam (R-DCB)}\label{subsec:R-DCB}
\subsubsection{Description of the R-DCB specimen}\label{subsec:R-DCB-Description}
\begin{figure}[h!]
	\centering
	\includegraphics[width=0.85\linewidth]{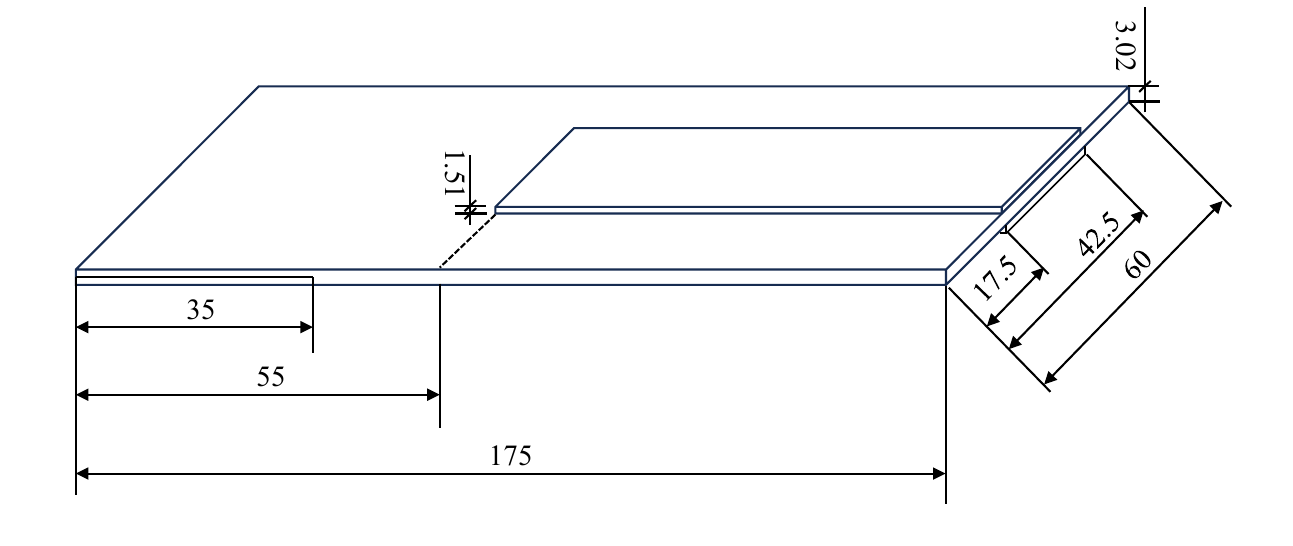}
	\caption{Geometry of the R-DCB specimen (units in mm)~\cite{carreras2019benchmark}}
	\label{fig:R-DCB-GEO}
\end{figure}

The R-DCB benchmark considered in this study is based on the experimental specimen proposed by Carreras et al.~\cite{carreras2019benchmark}. The specimen consists of a standard DCB laminate fabricated from 16 unidirectional plies oriented at $0^\circ$. Two reinforcement laminates, each composed of eight unidirectional plies with the same stacking sequence, are bonded to the upper and lower surfaces of the DCB laminate. The ply material properties are presented in Table~\ref{tab:R-DCB-Material}. The resin-rich layer is modelled by Araldite 2015, whose material properties are obtained from Campilho et al.~\cite{campilho2013modelling}. The corresponding resin-rich layer properties are summarized in Table~\ref{tab:Kvalue-RCB}. The geometry of the R-DCB specimen is shown in Figure~\ref{fig:R-DCB-GEO}.
\begin{table}[h!]
  \centering
  \caption{Material properties for the R-DCB specimen\cite{carreras2019simulation}}
    \begin{tabular}{lll}
    \toprule
 \multicolumn{3}{l}{\textbf{Unidirectional carbon fiber/epoxy
prepreg}}  \\
 \midrule
    $E_{11}$= 154 GPa & $E_{22}$= 8.5 GPa & $E_{33}$ = 8.5 GPa \\[0.3em]
    $\nu_{12}$ = 0.35 & $\nu_{13}$ = 0.35 & $\nu_{23}$ = 0.4  \\[0.3em]
    $G_{12}$ = 4.2 GPa & $G_{13}$ = 4.2 GPa & $G_{23}$ = 3.04 GPa \\[0.5em]
    Fracture toughness data & & \\[0.3em]
    $G_{\rm{Ic}} = 0.305 \, \mathrm{kJ/m^2} $ & $G_{\rm{IIc}} = 2.77 \, \mathrm{kJ/m^2}$ & $\eta = 2.05$ \\ [0.3em]
    Interfacial strength data \cite{ai2025structural} & & \\[0.3em]
    $\tau_{\rm{Ic}} = 32.6 \, \mathrm{MPa} $ & $\tau_{\rm{IIc}} = 98 \, \mathrm{MPa}$ &  \\ 
    \bottomrule
    \end{tabular}%
  \label{tab:R-DCB-Material}%
\end{table}%

\begin{table}[h!]
  \centering
  \caption{Material parameters of Araldite 2015~\cite{campilho2013modelling,grande1991effects}}
    \begin{tabular}{cccc}
    \toprule
    $E^{\textrm{R-DCB}}_{\textrm{rr}}$$\mathrm{(N/mm^2)}$ & $G^{\textrm{R-DCB}}_{\textrm{rr}}$$\mathrm{(N/mm^2)}$ & $\nu^{\textrm{R-DCB}}_\textrm{rr}$&  $h_{\textrm{rr}}$ $\mathrm{(mm)}$ \\[0.3em]
 \midrule
    1850 & 560 & 0.33 & 0.02286  \\[0.3em]
    \bottomrule
    \end{tabular}%
  \label{tab:Kvalue-RCB}%
\end{table}

\begin{figure}[h!]
	\centering
	\includegraphics[width=1.0\linewidth]{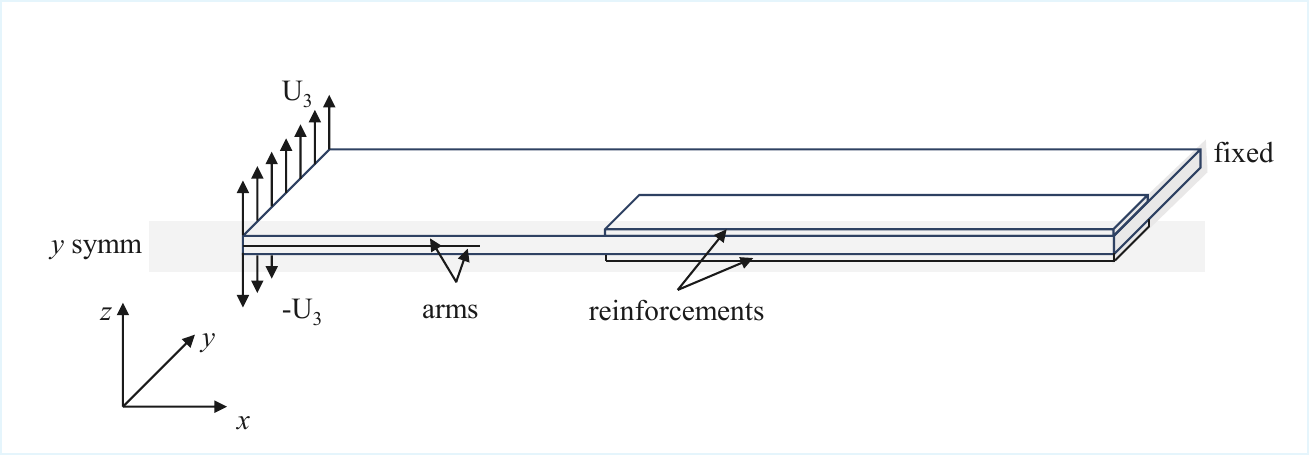}
	\caption{Boundary conditions and loading of the finite element model; U$_3$=7.5 mm}
	\label{fig:R-DCB-BC}
\end{figure}

Compared with the previously validated benchmark models, the R-DCB model has a larger geometric size and a higher number of finite elements. To improve computational efficiency, symmetry boundary conditions are employed, as illustrated in Figure~\ref{fig:R-DCB-BC}. The right end of the model is fixed, while a prescribed displacement is applied at the left end.

\subsubsection{Penalty stiffness of the unreinforced region in the R-DCB specimen}
\begin{figure}[h!]
	\centering
	\includegraphics[width=0.95\linewidth]{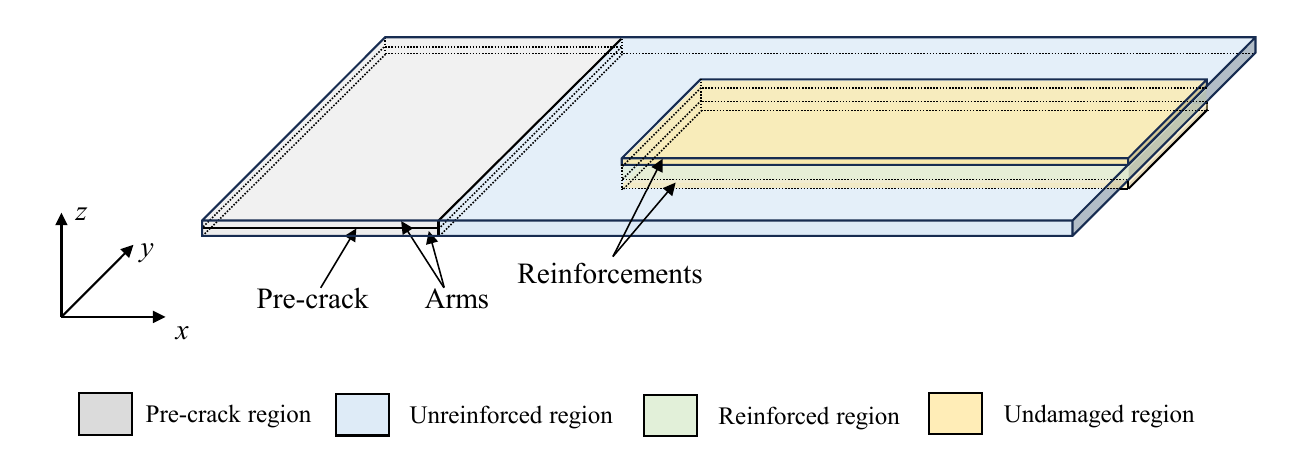}
	\caption{Different regions of the R-DCB specimen}
	\label{fig:Stress-R-DCB-region}
\end{figure}

\begin{figure}[h!]
	\centering
	\includegraphics[width=0.95\linewidth]{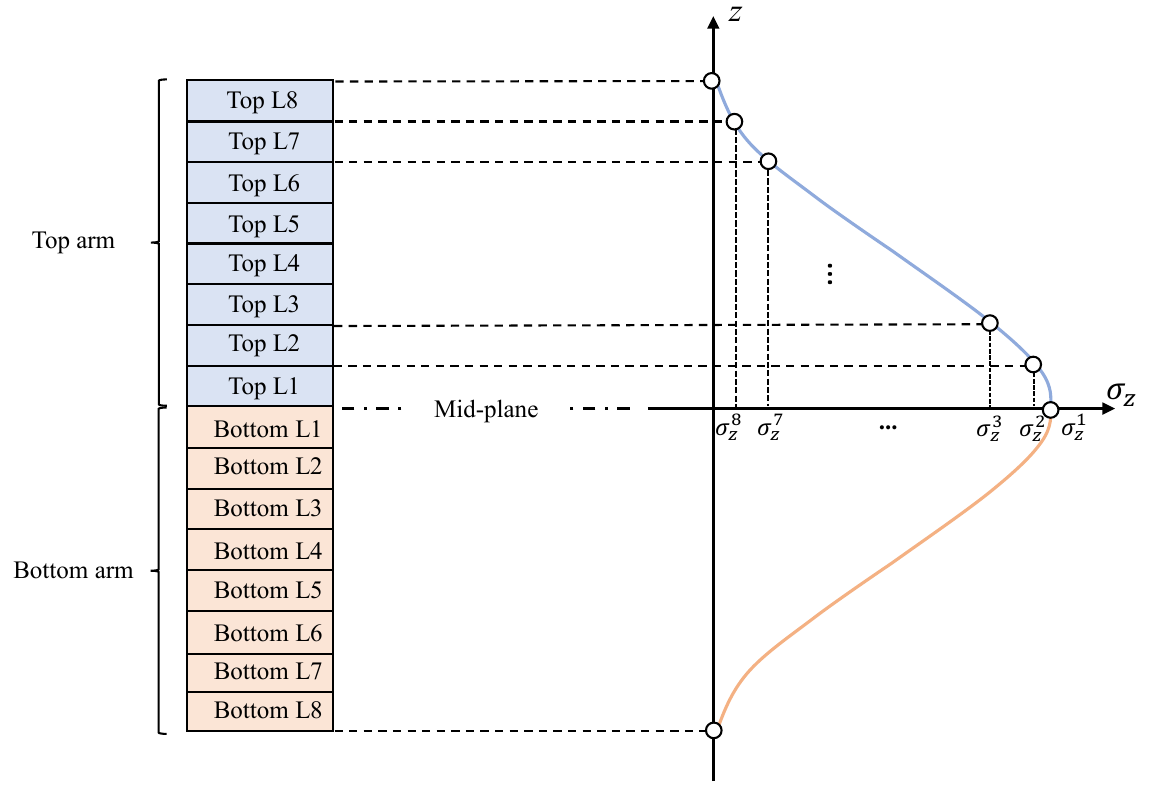}
	\caption{Through-thickness distribution of out-of-plane normal stress in the 16-ply unreinforced region of the R-DCB specimen}
	\label{fig:Stress-R-DCB-unreinforced}
\end{figure}

As shown in Figure~\ref{fig:Stress-R-DCB-region}, the R-DCB specimen can be divided into four regions: the pre-crack region, the unreinforced region, the reinforced region, and the undamaged region. Excluding the pre-crack region, structural cohesive elements are inserted along the mid-plane between the top and bottom laminate arms in both the unreinforced and reinforced regions to simulate crack initiation and propagation. During loading, these cohesive elements are progressively damaged as the crack propagates. Accordingly, the unreinforced and reinforced regions are collectively referred to as the damaged region, in contrast to the undamaged region, where no damage develops in the cohesive elements located between the laminate arms and the reinforcements. The unreinforced and reinforced regions each consist of a 16-ply laminate, whereas the undamaged region includes two additional 8-ply reinforcement laminates. The derivation of the stress distribution begins with the unreinforced region, as illustrated in Figure~\ref{fig:Stress-R-DCB-unreinforced}.

Based on the out-of-plane normal stress distribution shown in Figure~\ref{fig:Stress-R-DCB-unreinforced}, the ratio of the stress in each interface layer relative to the maximum stress value can be derived. The corresponding stress ratio values for each interface layer in the top arm are listed in Table~\ref{tab:Hrr-8Layer-normal}. 
\begin{table}[h!]
  \centering
  \caption{Out-of-plane normal stress ratios at the interface layers of the top arm in the 16-ply unreinforced region}
    \begin{tabular}{ccccccccc}
    \toprule
 $\dfrac{\sigma_z^1}{\sigma_z^{\textrm{max}}}$  & $\dfrac{\sigma_z^2}{\sigma_z^{\textrm{max}}}$ & $\dfrac{\sigma_z^3}{\sigma_z^{\textrm{max}}}$ &  $\dfrac{\sigma_z^4}{\sigma_z^{\textrm{max}}}$ &
 $\dfrac{\sigma_z^5}{\sigma_z^{\textrm{max}}}$&
 $\dfrac{\sigma_z^{\textrm{6}}}{\sigma_z^{\textrm{max}}}$ & $\dfrac{\sigma_z^7}{\sigma_z^{\textrm{max}}}$
  & $\dfrac{\sigma_z^8}{\sigma_z^{\textrm{max}}}$ 
    \\[0.8em]
 \midrule
   1.00 & 0.96 & 0.84 & 0.68 & 0.50 & 0.32 & 0.16 & 0.04 
 \\[0.3em]
    \bottomrule
    \end{tabular}%
  \label{tab:Hrr-8Layer-normal}%
\end{table}

Therefore, the sum of the out-of-plane stress ratios for the top arm in the unreinforced region can be calculated as:
\begin{align}\label{eq:total-hrr-normal-RDCB}
\sum_{i=1}^{n^{\textrm{Unreinforced}}_{\textrm{top}}}\, \frac{\sigma_z(z_i)}{\sigma_z^{\textrm{max}}} = \frac{\sigma_z^1+\sigma_z^2+\cdots+\sigma_z^8}{\sigma_z^{\textrm{max}}} = 4.5
\end{align}
where $n^{\textrm{Unreinforced}}_{\textrm{top}}$ is the number of plies in the top arm of the unreinforced region of the R-DCB specimen.

Due to the symmetric layup of the 16-layer unreinforced region, the sum of the ratios in Equation~\ref{eq:Kn_final} can be derived as follows:
\begin{align}
    \sum_{i=1}^{n^{\textrm{Unreinforced}}_{\textrm{top}}}\, \frac{\sigma_z(z_i)}{\sigma_z^{\textrm{max}}} +  \sum_{j=1}^{n^{\textrm{Unreinforced}}_{\textrm{bot}}}\, \frac{\sigma_z(z_j)}{\sigma_z^{\textrm{max}}} &=   2\sum_{i=1}^{8}\, \frac{\sigma_z(z_i)}{\sigma_z^{\textrm{max}}}  \nonumber\\[0.5em]
    &= 2\times4.5 =9
\end{align}
where $n^{\textrm{Unreinforced}}_{\textrm{bot}}$ is the number of plies in the bottom arm of the unreinforced region of the R-DCB specimen.

Based on Equation~\ref{eq:Kn_final}, the normal penalty stiffness for the unreinforced region in the R-DCB model can be calculated as follows:
\begin{align}\label{eq:Kn_R-DCB}
K^{\textrm{Unreinforced}}_{n}  &= \frac{1}
{\displaystyle\sum_{i=1}^{8}\, \frac{\sigma_z(z_i)}{\sigma_z^{\textrm{max}}}\, \frac{h_{\textrm{rr}}}{E^{\textrm{R-DCB}}_{\textrm{rr}}} +  \sum_{j=1}^{8}\, \frac{\sigma_z(z_j)}{\sigma_z^{\textrm{max}}}\,\frac{h_{\textrm{rr}}}{E^{\textrm{R-DCB}}_{\textrm{rr}}}} \nonumber\\[0.5em]
& = \frac{1}{9\times\frac{0.02286}{1850}} = 8992 \, \mathrm{N/mm^3} 
\end{align}

\begin{figure}[h!]
	\centering
	\includegraphics[width=0.95\linewidth]{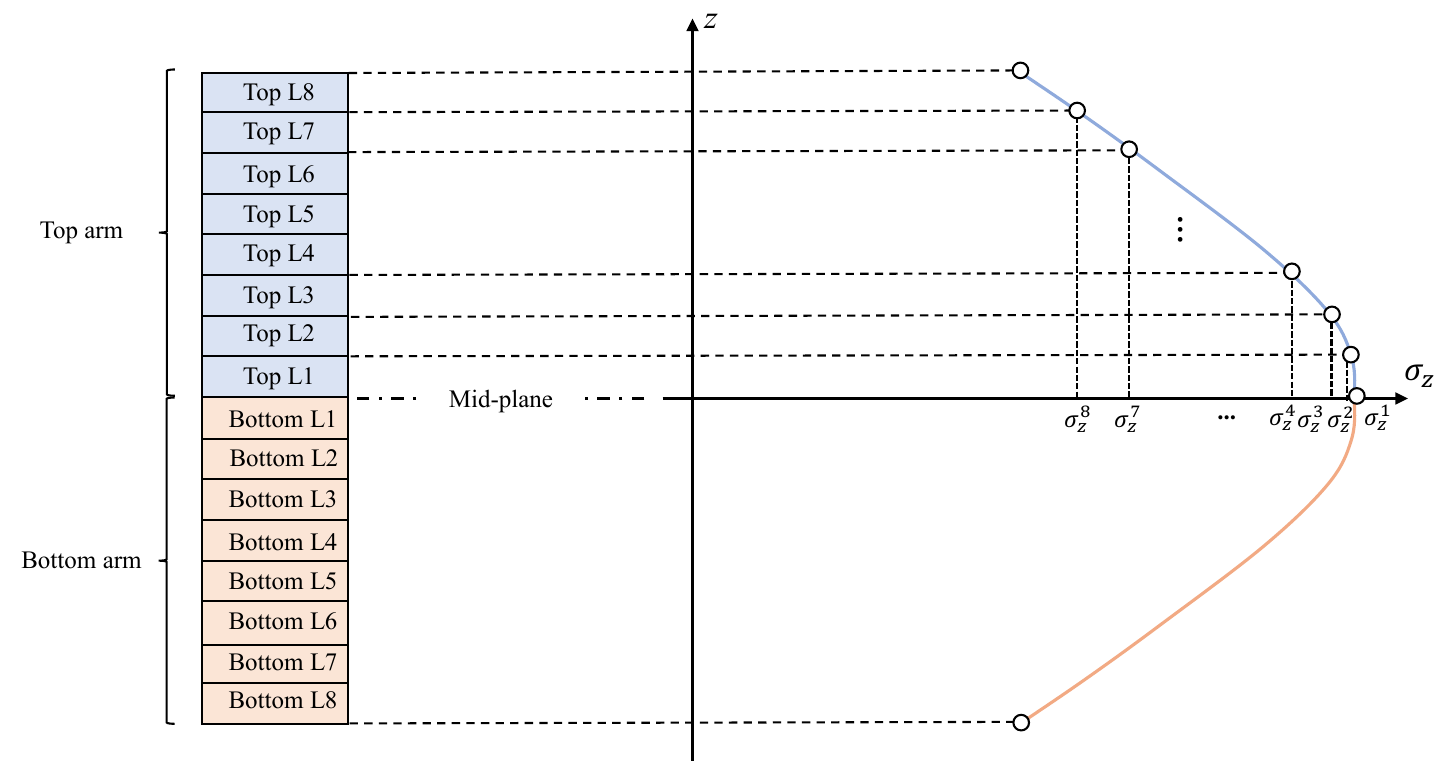}
	\caption{Through-thickness distribution of out-of-plane normal stress in the 16-ply reinforced region of the R-DCB specimen}
	\label{fig:Stress-R-DCB-reinforced}
\end{figure}

\subsubsection{Penalty stiffness of the reinforced region in the R-DCB specimen}
\label{subsubsec:reinforced-RDCB}
For the 16-ply reinforced region, the through-thickness distribution of the out-of-plane normal stress is shown in Figure~\ref{fig:Stress-R-DCB-reinforced}. Based on this distribution, interface stress ratios can be determined. The corresponding values for the interface layers in the top arm are listed in Table~\ref{tab:Hrr-16Layer-normal}. 
\begin{table}[h!]
  \centering
  \caption{Out-of-plane normal stress ratios at the interface layers of the top arm in the 16-ply reinforced region}
    \begin{tabular}{ccccccccc}
    \toprule
 $\dfrac{\sigma_z^1}{\sigma_z^{\textrm{max}}}$  & $\dfrac{\sigma_z^2}{\sigma_z^{\textrm{max}}}$ & $\dfrac{\sigma_z^3}{\sigma_z^{\textrm{max}}}$ &  $\dfrac{\sigma_z^4}{\sigma_z^{\textrm{max}}}$ &
 $\dfrac{\sigma_z^5}{\sigma_z^{\textrm{max}}}$&
 $\dfrac{\sigma_z^{\textrm{6}}}{\sigma_z^{\textrm{max}}}$ & $\dfrac{\sigma_z^7}{\sigma_z^{\textrm{max}}}$
  & $\dfrac{\sigma_z^8}{\sigma_z^{\textrm{max}}}$ 
    \\[0.8em]
 \midrule
   1.00 & 0.99 & 0.96 & 0.91 & 0.84 & 0.77 & 0.68 & 0.59 
\\[0.2em]
    \bottomrule
    \end{tabular}%
  \label{tab:Hrr-16Layer-normal}%
\end{table}

The sum of the out-of-plane normal stress ratios for the top arm in the reinforced region can be expressed as follows:
\begin{align}\label{eq:total-hrr-normal-RDCB-Reinforced}
\sum_{i=1}^{n^{\textrm{Reinforced}}_{\textrm{top}}}\, \frac{\sigma_z(z_i)}{\sigma_z^{\textrm{max}}} &= \frac{\sigma_z^1+\sigma_z^2+\cdots+\sigma_z^{8}}{\sigma_z^{\textrm{max}}}= 6.74
\end{align}
where $n^{\textrm{Reinforced}}_{\textrm{top}}$ is the number of plies in the top arm of the reinforced region of the R-DCB specimen.

Because the 16-ply reinforced region is symmetric about the laminate mid-plane, the sum of the out-of-plane normal stress ratios can be written as:
\begin{align}
    \sum_{i=1}^{n^{\textrm{Reinforced}}_{\textrm{top}}}\, \frac{\sigma_z(z_i)}{\sigma_z^{\textrm{max}}} +  \sum_{j=1}^{n^{\textrm{Reinforced}}_{\textrm{bot}}}\, \frac{\sigma_z(z_j)}{\sigma_z^{\textrm{max}}} &=   2\sum_{i=1}^{8}\, \frac{\sigma_z(z_i)}{\sigma_z^{\textrm{max}}} \nonumber\\[0.5em]
    &= 2\times6.74 =13.48
\end{align}
where $n^{\textrm{Reinforced}}_{\textrm{bot}}$ is the number of plies in the bottom arm of the reinforced region of the R-DCB specimen.

Accordingly, the normal penalty stiffness for the reinforced region in the R-DCB specimen is given by:
\begin{align}\label{eq:Kn_R-DCB-reinforced}
K^{\textrm{Reinforced}}_{n}  &= \frac{1}
{\displaystyle\sum_{i=1}^{8}\, \frac{\sigma_z(z_i)}{\sigma_z^{\textrm{max}}}\, \frac{h_{\textrm{rr}}}{E^{\textrm{R-DCB}}_{\textrm{rr}}} +  \sum_{j=1}^{8}\, \frac{\sigma_z(z_j)}{\sigma_z^{\textrm{max}}}\,\frac{h_{\textrm{rr}}}{E^{\textrm{R-DCB}}_{\textrm{rr}}}} \nonumber\\[0.5em]
& = \frac{1}{13.48\times\frac{0.02286}{1850}} = 6004 \, \mathrm{N/mm^3} 
\end{align}

The R-DCB model is similar to the DCB model and exhibits Mode I fracture behavior. Therefore, the effect of the shear penalty stiffness of the damaged region is negligible and is not considered further in the analyses of either the unreinforced or reinforced region.

\subsubsection{Penalty stiffness of the undamaged region in the R-DCB specimen}
\begin{figure}[h!]
	\centering
	\includegraphics[width=0.9\linewidth]{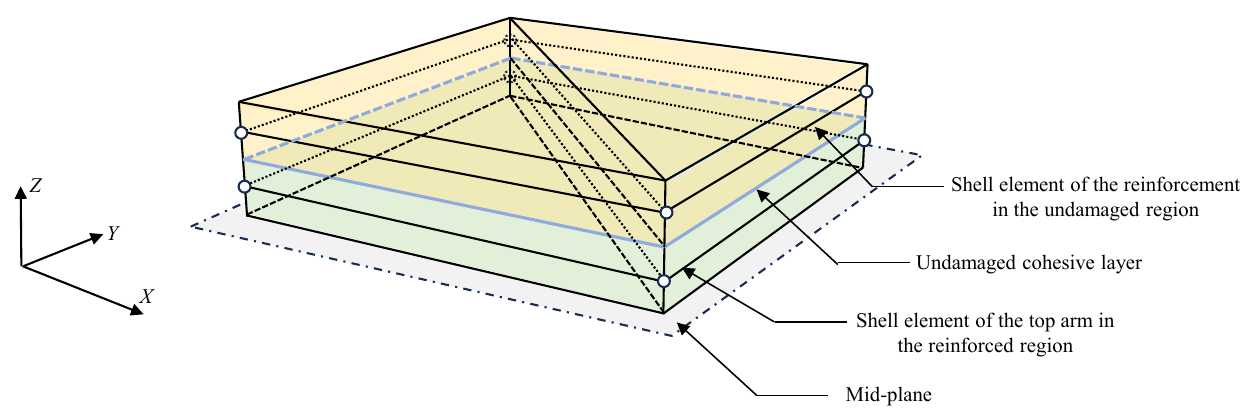}
	\caption{Finite element model of the undamaged region in the R-DCB specimen}
	\label{fig:Stress-R-DCB-undamaged-region}
\end{figure}

Figure~\ref{fig:Stress-R-DCB-undamaged-region} illustrates the undamaged region of the R-DCB specimen. The top arm and the reinforcement are modelled by triangular higher-order shell elements, while an undamaged structural cohesive layer is inserted between them to represent the bonding interface. Although no damage develops in this region throughout the loading process, the structural cohesive elements are retained to ensure displacement continuity and to transfer interlaminar tractions. Unlike the DCB benchmark, the shear penalty stiffness can no longer be neglected and must be determined from the through-thickness transverse shear stress distribution. The derivation of the corresponding normal and shear penalty stiffnesses is presented below.

\begin{figure}[h!]
	\centering
	\includegraphics[width=0.98\linewidth]{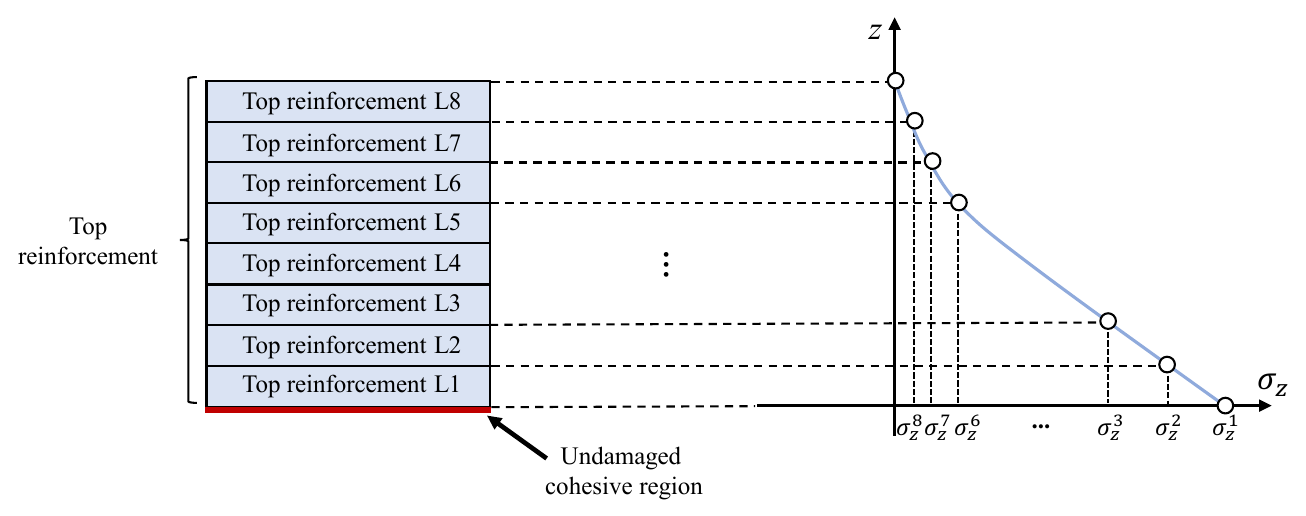}
	\caption{Through-thickness distribution of out-of-plane normal stress in the 8-ply undamaged region of the R-DCB specimen}
	\label{fig:Stress-R-DCB-undamaged-normal}
\end{figure}

\begin{figure}[h!]
	\centering
	\includegraphics[width=0.98\linewidth]{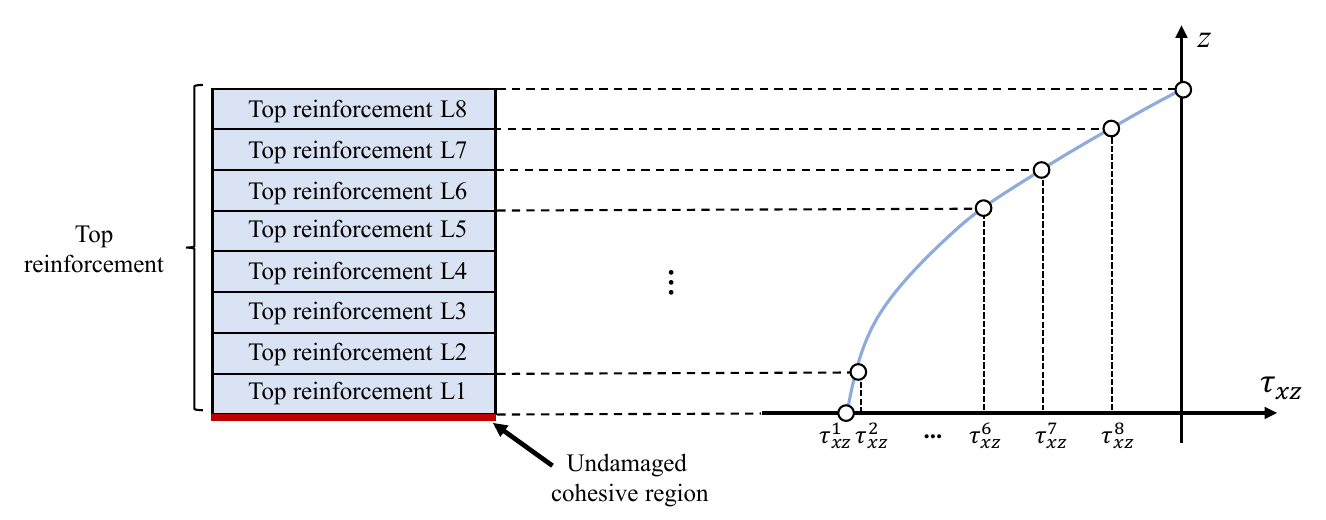}
	\caption{Through-thickness distribution of transverse shear stress in the 8-ply undamaged region of the R-DCB specimen}
	\label{fig:Stress-R-DCB-undamaged-shear}
\end{figure}

As shown in Figure~\ref{fig:Stress-R-DCB-undamaged-region}, only the through-thickness stress distribution in the reinforcement is required to derive the penalty stiffness for the undamaged region. Because the R-DCB specimen is symmetric about the mid-plane, the penalty stiffnesses of the undamaged regions in the top and bottom reinforcements are identical. Therefore, the following derivation is presented for the top arm only. The corresponding distributions of the out-of-plane normal stress and transverse shear stress are shown in Figure~\ref{fig:Stress-R-DCB-undamaged-normal} and \ref{fig:Stress-R-DCB-undamaged-shear}, respectively.

Based on the through-thickness distributions of the out-of-plane normal and transverse shear stresses shown in Figures~\ref{fig:Stress-R-DCB-undamaged-normal} and~\ref{fig:Stress-R-DCB-undamaged-shear}, the interface stress ratios relative to the corresponding maximum values are determined. The resulting normal and shear stress ratios for each interface layer are presented in Tables~\ref{tab:Hrr-8Layer-normal-undamaged} and \ref{tab:Hrr-8Layer-shear-undamaged}, respectively.

\begin{table}[h!]
  \centering
  \caption{Out-of-plane normal stress ratios at the interface layers of the 8-ply undamaged region}
    \begin{tabular}{ccccccccc}
    \toprule
 $\dfrac{\sigma_{z}^1}{\sigma_{z}^{\textrm{max}}}$  & $\dfrac{\sigma_{z}^2}{\sigma_{z}^{\textrm{max}}}$ & $\dfrac{\sigma_{z}^3}{\sigma_{z}^{\textrm{max}}}$ &  $\dfrac{\sigma_{z}^4}{\sigma_{z}^{\textrm{max}}}$ &
 $\dfrac{\sigma_{z}^5}{\sigma_{z}^{\textrm{max}}}$&
 $\dfrac{\sigma_{z}^{\textrm{6}}}{\sigma_{z}^{\textrm{max}}}$ & $\dfrac{\sigma_{z}^7}{\sigma_{z}^{\textrm{max}}}$
  & $\dfrac{\sigma_{z}^8}{\sigma_{z}^{\textrm{max}}}$ 
    \\[0.8em]
 \midrule
   1.00 & 0.81 & 0.63 & 0.46 & 0.31 & 0.18 & 0.09 & 0.02
 \\[0.2em]
    \bottomrule
    \end{tabular}%
  \label{tab:Hrr-8Layer-normal-undamaged}%
\end{table}

\begin{table}[h!]
  \centering
  \caption{Transverse shear stress ratios at the interface layers of the 8-ply undamaged region}
    \begin{tabular}{ccccccccc}
    \toprule
 $\dfrac{\tau_{xz}^1}{\tau_{xz}^{\textrm{max}}}$  & $\dfrac{\tau_{xz}^2}{\tau_{xz}^{\textrm{max}}}$ & $\dfrac{\tau_{xz}^3}{\tau_{xz}^{\textrm{max}}}$ &  $\dfrac{\tau_{xz}^4}{\tau_{xz}^{\textrm{max}}}$ &
 $\dfrac{\tau_{xz}^5}{\tau_{xz}^{\textrm{max}}}$&
 $\dfrac{\tau_{xz}^{\textrm{6}}}{\tau_{xz}^{\textrm{max}}}$ & $\dfrac{\tau_{xz}^7}{\tau_{xz}^{\textrm{max}}}$
  & $\dfrac{\tau_{xz}^8}{\tau_{xz}^{\textrm{max}}}$ 
    \\[0.8em]
 \midrule
     1.00 & 0.98 & 0.94 & 0.86 & 0.75 & 0.61 & 0.44 & 0.23 
 \\[0.2em]
    \bottomrule
    \end{tabular}%
  \label{tab:Hrr-8Layer-shear-undamaged}%
\end{table}

The sums of the out-of-plane normal stress ratios and transverse shear stress ratios in the undamaged region are given by:
\begin{align}\label{eq:R-DCB-undamaged-stress-sum-normal}
\sum_{i=1}^{n^{\textrm{Undamaged}}}\, \frac{\sigma_{z}(z_i)}{\sigma_{z}^{\textrm{max}}} &= \frac{\sigma_{z}^1+\sigma_{z}^2+\cdots+\sigma_{z}^{8}}{\sigma_{z}^{\textrm{max}}}= 3.5
\end{align}
and
\begin{align}\label{eq:R-DCB-undamaged-stress-sum-shear}
\sum_{i=1}^{n^{\textrm{Undamaged}}}\, \frac{\tau_{xz}(z_i)}{\tau_{xz}^{\textrm{max}}} &= \frac{\tau_{xz}^1+\tau_{xz}^2+\cdots+\tau_{xz}^{8}}{\tau_{xz}^{\textrm{max}}}= 5.81
\end{align}
where $n^{\textrm{Undamaged}}$ is the number of plies in the undamaged region of the R-DCB specimen.

Substituting these stress ratios into Equations~\ref{eq:Kn_final} and \ref{eq:Ks_final} yields the normal and shear penalty stiffnesses for the undamaged region of the R-DCB specimen, respectively, as follows:
\begin{align}\label{eq:Kn_Undamaged}
    K^{\textrm{Undamaged}}_{n}  &= \frac{1}
{\displaystyle\sum_{i=1}^{16}\, \frac{\sigma_z(z_i)}{\sigma_z^{\textrm{max}}}\, \frac{h_{\textrm{rr}}}{E^{\textrm{R-DCB}}_{\textrm{rr}}}} \nonumber\\[0.5em]
& = \frac{1}{3.5\times\frac{0.02286}{1850}} = 23122 \, \mathrm{N/mm^3}  \\[0.5em]
K^{\textrm{Undamaged}}_s  &= \frac{1}
{\displaystyle\sum_{i=1}^{16}\, \frac{\tau_{xz}(z_i)}{\tau_{xz}^{\textrm{max}}}\, \frac{h_{\textrm{rr}}}{G^{\textrm{R-DCB}}_{\textrm{rr}}}} \nonumber\\[0.5em]
& = \frac{1}{5.81\times\frac{0.02286}{560}} = 4216 \, \mathrm{N/mm^3} 
\end{align}

\subsubsection{Load-displacement curves}\label{subsubsec:RF-U-RDCB}
\begin{figure}[h!]
      \centering
	   \begin{subfigure}{0.48\linewidth}
		\includegraphics[scale=1]{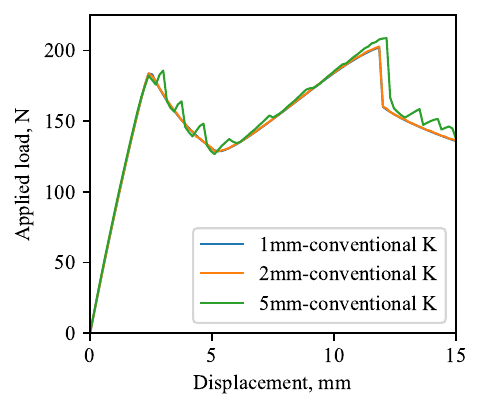}
		\caption{Conventional penalty stiffness}
		\label{fig:RDCB-comparison-Turon}
	   \end{subfigure}
	   \begin{subfigure}{0.48\linewidth}
		\includegraphics[scale=1]{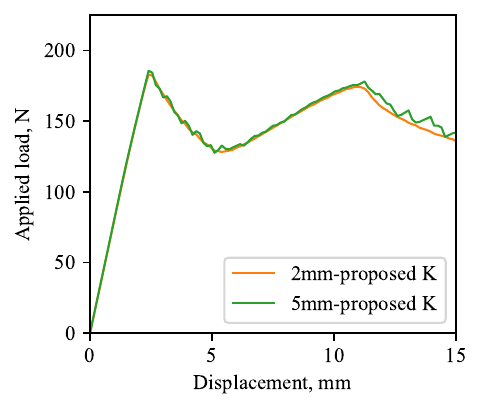}
		\caption{Proposed penalty stiffness}
		\label{fig:RDCB-comparison-XAI}
	    \end{subfigure}
	\caption{Comparison of the load-displacement curves obtained from the R-DCB simulations using the conventional and proposed penalty stiffness formulations at different mesh densities}
	\label{fig:RDCB-comparison}
\end{figure}

The load-displacement curves obtained from the R-DCB simulations are shown in Figure~\ref{fig:RDCB-comparison}. The left panel (Figure~\ref{fig:RDCB-comparison-Turon}) shows the results obtained by the conventional penalty stiffness, whereas the right panel (Figure~\ref{fig:RDCB-comparison-XAI}) presents the results obtained by the proposed penalty stiffness.

A comparison of the results in Figures~\ref{fig:RDCB-comparison-Turon} and~\ref{fig:RDCB-comparison-XAI} shows a clear difference in mesh convergence. When the conventional penalty stiffness is adopted, the load-displacement curve obtained with a 5 mm mesh differs noticeably from those obtained with the 2 mm and 1 mm meshes. In particular, the second peak occurs at a larger displacement and reaches a higher load, and the descending segment exhibits obvious oscillations. By contrast, the results obtained with the 2 mm and 1 mm meshes are in close agreement. When the proposed penalty stiffness is adopted, the results obtained with a 5 mm mesh closely matches that of the 2 mm mesh, and the oscillations are significantly reduced. These results demonstrate that the proposed penalty stiffness improves the mesh convergence.

In addition to the improved mesh convergence, a further difference can be observed in the response around the second load peak. As shown in Figure~\ref{fig:RDCB-comparison-Turon}, the load drops abruptly after reaching the second peak with the conventional penalty stiffness. By contrast, the proposed penalty stiffness (Figure~\ref{fig:RDCB-comparison-XAI}) produces a more gradual softening process, which results in a second peak load 14\% lower than that predicted with the conventional penalty stiffness.
\begin{figure}[h!]
	\centering
	\includegraphics[scale=1]{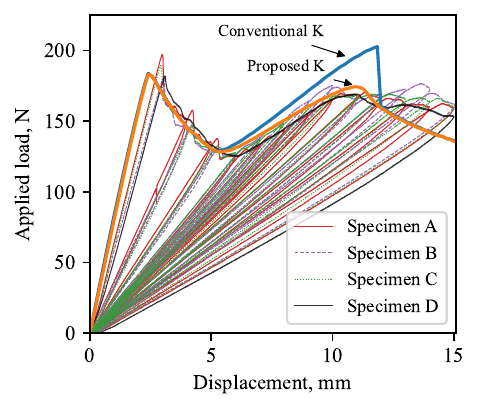}
	\caption{Comparison of numerical and experimental results of the load-displacement curves~\cite{carreras2019benchmark}}
	\label{fig:RDCB-Comparison-2mm}
\end{figure}

To assess the accuracy of the two penalty stiffness formulations, the load-displacement curves obtained using a 2 mm mesh are compared with the experiment data, as shown in Figure~\ref{fig:RDCB-Comparison-2mm}. The results show that the curve obtained using the proposed penalty stiffness agrees more closely with the experimental results. In particular, the second loading segment exhibits a lower stiffness than that predicted using the conventional penalty stiffness. Consequently, the predicted second peak load is in much better agreement with the experimental results. This improvement can be attributed to the proposed penalty stiffness formulation, which incorporates the transverse shear compliance. As a result, the proposed method avoids an excessively stiff response when the reinforced region is loaded.

In summary, for complex composite structures with reinforcements or stiffeners, the proposed penalty stiffness provides improved mesh convergence and higher predictive accuracy. Consequently, it enables the use of coarser meshes without losing accuracy, thereby improving the computational efficiency of delamination simulations.

\subsubsection{Crack front evolution}\label{subsubsec:Damage map}
\begin{figure}[h!]
	\centering
		 \includegraphics[scale=1]{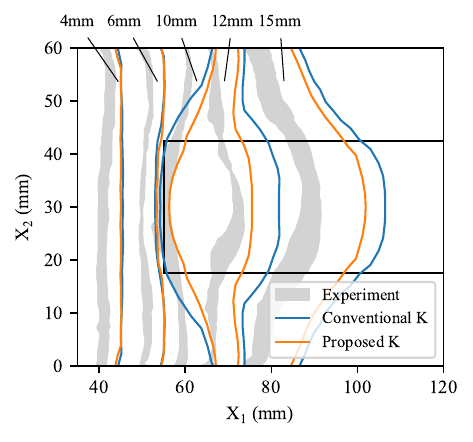}
		 \caption{Comparison of numerical and experimental~\cite{carreras2019benchmark} delamination front positions}
		 \label{fig:R-DCB-Damage}
\end{figure}

Figure~\ref{fig:R-DCB-Damage} compares the crack front positions obtained from the experiments and numerical simulations at different opening displacements. Regardless of whether the conventional or proposed penalty stiffness is adopted, the predicted crack fronts are in good agreement with the experimental results. At opening displacements below 6 mm, the crack front remains approximately straight across the specimen width. As the opening displacement exceeds 6 mm, it gradually evolves into a curved profile. At an opening displacement of 10 mm, the central portion of the crack front begins to lag behind the free edges, which indicates a lower crack propagation rate in the central region. As the opening displacement increases further, the crack front continues to evolve. The propagation rate in the central region eventually exceeds that at the free edges, resulting in a convex crack front in the direction of crack propagation.

Overall, the crack front evolution predicted by the two penalty stiffness formulations is generally similar. Before the crack reaches the reinforced region (i.e., at opening displacements of 6 mm or less), both predictions agree well with the experimental observations. Once the crack propagates into the reinforced region, obvious differences become evident. At an opening displacement of 10 mm, the crack front predicted using the proposed penalty stiffness extends further into the reinforced region than that predicted using the conventional penalty stiffness. This prediction shows better agreement with the experimental results. As the opening displacement increases beyond 10 mm, the crack propagation trend reverses. The crack propagation rate predicted using the proposed penalty stiffness becomes lower than that predicted using the conventional penalty stiffness. Consequently, the predicted crack length is approximately 14\% shorter and agrees more closely with the experimental results.

Both numerical models slightly overpredict crack propagation because the compliance of the testing machine is not included in the numerical model. Nevertheless, the proposed penalty stiffness provides a closer match to the experimental observations throughout the crack propagation process. These results indicate that the proposed formulation improves the accuracy of crack propagation predictions.

\section{Summary and conclusions}
\label{sec:summary}
This work aims at (a) deriving distributions of the out-of-plane normal and transverse shear stresses, and (b) establishing the accurate penalty stiffness formulation directly from the material properties and the thickness of the resin-rich layer by incorporating these stress distributions. In the previous formulation, the penalty stiffness is determined solely from the resin-rich layer thickness without considering stress distributions. Consequently, it requires a layer-by-layer representation of the laminate and cannot be directly applied to equivalent single-layer models. As a result, the computational efficiency is significantly reduced. The proposed method is based on beam theory and derives analytical expressions for the out-of-plane normal and transverse shear stress distributions through the laminate thickness. The opening displacement is then determined from the ratio of the stress at each interface layer to the maximum stress. The corresponding penalty stiffness is then calculated from the opening displacement. For mixed-mode fracture problems, a mode-dependent cohesive law is employed to accurately predict damage evolution.

The proposed method has been implemented in three-dimensional structural-element~\cite{ai2025structural} models and validated using DCB, ENF, MMB, and R-DCB unidirectional laminate benchmarks. By overcoming limitations of the structural cohesive element implementations in equivalent single-layer models, the proposed approach significantly improves computational efficiency. Comparisons with the conventional penalty stiffness show that the proposed penalty stiffness substantially improves mesh convergence and prediction accuracy. Consequently, it allows cohesive elements to be discretized with coarser meshes, thereby reducing computational time. An analysis of the crack-tip stress distribution in the DCB model shows that the proposed method eliminates the mesh dependency observed in previous high-order structural elements with respect to compressive stress distribution ahead of the crack tip. In addition, it reduces the discrepancy between structural element and solid element predictions. For fully three-dimensional delamination problems with highly curved crack fronts, the proposed method also provides better agreement with experimental results than existing approaches.

The proposed method enables accurate prediction of out-of-plane delamination and the corresponding out-of-plane stress distributions in three-dimensional composite laminate models based on the equivalent single-layer method. Future work will extend the proposed formulation to predict in-plane damage mechanisms, including matrix cracking and fiber failure. More complex failure modes will be investigated by developing a three-dimensional damage criterion that combines the in-plane stresses obtained from shell elements with the out-of-plane stresses provided by cohesive elements.

\section*{Acknowledgements}
The first author would like to acknowledge funding support from the China Scholarship Council (No.201906290034) for this research.


\bibliographystyle{elsarticle-num-names} 
\bibliography{cas-refs}







\end{document}